\documentclass{elsart5p}

\usepackage{natbib}

\usepackage{graphicx}
\usepackage{color}

\usepackage{amssymb}
\usepackage{amsmath}
\usepackage{listings}
\newcommand{\itbold}[1]{\textbf{\textit{#1}}}

\newcommand{\bm}[1]{\boldsymbol  #1 }

\begin{document}

\begin{frontmatter}



\title{Phantom-GRAPE: numerical software library to accelerate
  collisionless $N$-body simulation with SIMD instruction set on x86
  architecture}


\author[UTsukuba]{Ataru Tanikawa\corauthref{cor1}},
\ead{tanikawa@ccs.tsukuba.ac.jp}
\author[UTsukuba]{Kohji Yoshikawa},
\author[UTsukuba]{Keigo Nitadori} \&
\author[UTsukuba]{Takashi Okamoto}

\address[UTsukuba]{Center for Computational Science, University of
  Tsukuba, 1--1--1, Tennodai, Tsukuba, Ibaraki 305--8577, Japan}

\corauth[cor1]{Corresponding author.}

\begin{abstract}
 We have developed a numerical software library for collisionless
 $N$-body simulations named ``Phantom-GRAPE'' which highly accelerates
 force calculations among particles by use of a new SIMD instruction
 set extension to the x86 architecture, Advanced Vector eXtensions
 (AVX), an enhanced version of the Streaming SIMD Extensions (SSE). In
 our library, not only the Newton's forces, but also central forces
 with an arbitrary shape $f(r)$, which has a finite cutoff radius
 $r_{\rm cut}$ (i.e. $f(r)=0$ at $r>r_{\rm cut}$), can be quickly
 computed. In computing such central forces with an arbitrary force
 shape $f(r)$, we refer to a pre-calculated look-up table. We also
 present a new scheme to create the look-up table whose binning is
 optimal to keep good accuracy in computing forces and whose size is
 small enough to avoid cache misses. Using an Intel Core i7--2600
 processor, we measure the performance of our library for both of the
 Newton's forces and the arbitrarily shaped central forces. In the
 case of Newton's forces, we achieve $2 \times 10^9$ interactions per
 second with one processor core (or $75$~GFLOPS if we count $38$
 operations per interaction), which is $20$ times higher than the
 performance of an implementation without any explicit use of SIMD
 instructions, and $2$ times than that with the SSE instructions.
 With four processor cores, we obtain the performance of $8 \times
 10^9$ interactions per second (or $300$~GFLOPS). In the case of the
 arbitrarily shaped central forces, we can calculate $1 \times 10^9$
 and $4 \times 10^9$ interactions per second with one and four
 processor cores, respectively. The performance with one processor
 core is $6$ times and $2$ times higher than those of the
 implementations without any use of SIMD instructions and with the SSE
 instructions. These performances depend only weakly on the number of
 particles, irrespective of the force shape. It is good contrast with
 the fact that the performance of force calculations accelerated by
 graphics processing units (GPUs) depends strongly on the number of
 particles. Substantially weak dependence of the performance on the
 number of particles is suitable to collisionless $N$-body
 simulations, since these simulations are usually performed with
 sophisticated $N$-body solvers such as Tree- and TreePM-methods
 combined with an individual timestep scheme. We conclude that
 collisionless $N$-body simulations accelerated with our library have
 significant advantage over those accelerated by GPUs, especially on
 massively parallel environments.
\end{abstract}


\end{frontmatter}

\section{Introduction}

Self-gravity is one of the most essential physical processes in the
universe, and plays important roles in almost all categories of
astronomical objects such as globular clusters, galaxies, galaxy
clusters, etc. In order to follow the evolution of such systems,
gravitational $N$-body solvers have been widely used in numerical
astrophysics.

Due to prohibitively expensive computational cost in directly solving
$N$-body problems, many efforts have been made to reduce it in various
ways.  For example, several sophisticated algorithms to compute
gravitational forces among many particles with reduced computational
cost have been developed, such as Tree method \citep{Barnes86}, PPPM
method \citep{Hockney81}, TreePM method \citep{Xu95}, etc.

Another approach is to improve the computational performance with the
aid of additional hardware, such as GRAPE (GRAvity PipE) systems,
special-purpose accelerators for gravitational $N$-body simulations
\citep{Sugimoto90, Makino03, Fukushige05}, and general-purpose
computing on Graphics Processing Units (GPGPUs). GRAPE systems have
been used for further improvement of existing $N$-body solvers such as
Tree method \citep{Makino91}, PPPM method \citep{Brieu95,Yoshikawa05},
TreePM method \citep{Yoshikawa05}, P$^2$M$^2$ tree method
\citep{Kawai04}, and PPPT method \citep{Oshino11}. They have also
adapted to simulation codes for dense stellar systems based on
fourth-order Hermite scheme, such as {\tt NBODY4} \citep{Johnson06},
{\tt NBODY1} \citep{Harfst07}, {\tt kira} \citep{Simon08}, and {\tt
  GORILLA} \citep{Tanikawa09}. Recently, \citet{Hamada07},
\citet{Simon07}, \citet{Gaburov09}, and \citet{Bedorf12} explored the
capability of commodity graphics processing units (GPUs) as hardware
accelerators for $N$-body simulations and achieved similar to or even
higher performance than the GRAPE-6A and GRAPE-DR board.

A different approach to improve the performance of $N$-body
calculations is to utilize Streaming SIMD Extensions (hereafter SSE),
a SIMD (Single Instruction, Multiple Data) instruction set implemented
on x86 and x86\_64 processors. \citet{Nitadori06} exploited the SSE
and SSE2 instruction sets, and achieved speeding up of the Hermite
scheme \citep{Makino92} in mixed precision for collisional
self-gravitating systems. Although unpublished in literature,
Nitadori, Yoshikawa, \& Makino have also developed a numerical library
for $N$-body calculations in single-precision for collisionless
self-gravitating systems in which two-body relaxation is not
physically important and therefore single-precision floating-point
arithmetic suffices for the required numerical accuracy. Furthermore,
along this approach, they have also improved the performance in
computing arbitrarily-shaped forces with a cutoff distance, defined by
a user-specified function of inter-particle separation.  Such
capability to compute force shapes other than Newton's inverse-square
gravity is necessary in PPPM, TreePM, and Ewald methods. It should be
noted that GRAPE-5 and the later families of GRAPE systems have
similar capability to compute the Newton's force multiplied by a
user-specified cutoff function \citep{Kawai00}, and can be used to
accelerate PPPM and TreePM methods for cosmological $N$-body
simulations \citep{Yoshikawa05}. Based on these achievements, a
publicly available software package to improve the performance of both
collisional and collisionless $N$-body simulations has been developed,
which was named ``Phantom-GRAPE'' after the conventional GRAPE
system. A set of application programming interfaces of Phantom-GRAPE
for collisionless simulations is compatible to that of
GRAPE-5. Phantom-GRAPE is widely used in various numerical simulations
for galaxy formation \citep{Saitoh08,Saitoh09} and the cosmological
large-scale structures
\citep{Ishiyama08,Ishiyama09a,Ishiyama09b,Ishiyama10,Ishiyama11}.

Recently, a new processor family with ``Sandy Bridge''
micro-architecture\footnote{\ttfamily
  http://www.intel.com/content/dam/doc/manual/} by Intel Corporation
and that with ``Bulldozer'' micro-architecture\footnote{\ttfamily
  http://support.amd.com/us/Processor\_TechDocs/} by AMD Corporation
have been released. Both of the processors support a new set of
instructions known as Advanced Vector eXtensions (AVX), an enhanced
version of the SSE instructions. In the AVX instruction set, the width
of the SIMD registers is extended from 128-bit to 256-bit. We can
perform SIMD operations on two times larger data than
before. Therefore, the performance of a calculation with the AVX
instructions should be two times higher than that with the SSE
instructions if the execution unit is also extended to 256-bit.

\citet{Tanikawa11} (hereafter, paper I) developed a software library
for {\it collisional} $N$-body simulations using the AVX instruction
set in the mixed precision, and achieved a fairly high performance. In
this paper, we present a similar library implemented with the AVX
instruction set but for {\it collisionless} $N$-body simulations in
single-precision.

The structure of this paper is as follows. In section \ref{sec:avx},
we overview the AVX instruction set. In section
\ref{sec:implementation}, we describe the implementation of
Phantom-GRAPE. In section \ref{sec:accuracy} and
\ref{sec:performance}, we show the accuracy and performance,
respectively. In section \ref{sec:summary}, we summarize this paper.

\section{The AVX instruction set}
\label{sec:avx}

In this section, we present a brief review of the Advanced Vector
eXtensions (AVX) instruction set. Details of the difference between
SSE and AVX is described in section 3.1 of paper I. AVX is a SIMD
instruction set as well as SSE, and supports many operations, such as
addition, subtraction, multiplication, division, square-root,
approximate inverse-square-root, several bitwise operations, etc. In
such operations, dedicated registers with 256-bit length called ``YMM
registers'' are used to store the eight single-precision
floating-point numbers or four double-precision floating-point
numbers.  Note that the lower 128-bit of the YMM registers have alias
name ``XMM registers'', and can be used as the dedicated registers for
the SSE instructions for a backward compatibility.

An important feature of AVX and SSE instruction sets is the fact that
they have a special instruction for a very fast approximation of
inverse-square-root with an accuracy of about 12-bit. Actually, this
instruction is quite essential to improve the performance of the
gravitational force calculations, since the most expensive part in the
force calculation is an execution of inverse-square-root of squared
distances of the particle pairs. As already discussed in
\citet{Nitadori06}, the approximate values can be adopted as initial
values of the Newton-Raphson iteration to improve the accuracy, and we
can obtain 24-bit accuracy after one Newton-Raphson iteration. For
collisionless self-gravitating systems, however, the accuracy of
$\simeq$ 12 bits is sufficient because the accuracy of
inverse-square-root does not affect the resultant force accuracy if
one adopts an approximate $N$-body solver such as Tree, PPPM and
TreePM methods. Therefore, we use the raw approximate instruction
throughout this study.

Since the present-day compilers cannot always detect concurrency of
the loops effectively, and cannot fully resolve the mutual dependency
among data in the code, it is quite rare that compilers generate codes
with SIMD instructions in effective manners from codes expressed in
high-level languages. For an efficient use of the AVX instructions,
we need to program with assembly-languages explicitly or
compiler-dependent intrinsic functions and data type extensions. In
assembly-languages, we can manually control the assignment of YMM
registers to computational data, and minimize the access to the main
memory by optimizing the assignment of each register. In this work, we
adopt an implementation of the AVX instructions using inline-assembly
language with C expression operands, embedded in C-language, which is
a part of language extensions of GCC (GNU Compiler Collection).

\section{Implementation}
\label{sec:implementation}

Here, we describe the detailed implementation to accelerate $N$-body
calculation using the AVX instructions. For a given set of positions
$\itbold{r}_i$ of $N$ particles, we try to accelerate the calculations
of a gravitational force given as follows:
\begin{equation}
  \label{eq:newton_force}
  \itbold{a}_i = \sum_{j=1}^N \frac{Gm_j (\itbold{r}_j -
    \itbold{r}_i)}{(|\itbold{r}_j - \itbold{r}_i|^2
    +\epsilon^2)^{3/2}},
\end{equation}
where $G$ is the gravitational constant, $m_j$ the mass of the $j$-th
particle, and $\epsilon$ the gravitational softening length.  In
addition to that, we also try to accelerate the computations of
central forces among particles with an arbitrary force shape $f(r)$
given by
\begin{equation}
  \label{eq:arbitrary_force}
  \itbold{a}_i=\sum_{j=1}^N m_j f(|\itbold{r}_j-\itbold{r}_i|)\frac{\itbold{r}_j-\itbold{r}_i}{|\itbold{r}_j-\itbold{r}_i|}, 
\end{equation}
where $f(r)$ specifies the shape of the force as a function of
inter-particle separation $r$ with a cutoff distance $r_{\rm cut}$
(i.e. $f(r)=0$ at $r>r_{\rm cut}$). In the above expressions,
particles with subscript ``$j$'' exert forces on those with subscript
``$i$''.  In the rest of this paper, the former are referred to as
``$j$-particles'', and the latter as ``$i$-particles'' just for
convenience.

Since individual forces exerted by $j$-particles on $i$-particles can
be computed independently, we can calculate forces exerted by multiple
$j$-particles on multiple $i$-particles in parallel. As described in
the previous section, the AVX instructions can execute operations of
eight single-precision floating-point numbers on YMM registers in
parallel. By utilizing this feature of the AVX instructions, the
forces on four $i$-particles from two $j$-particles can be computed
simultaneously in a SIMD manner.

\begin{figure*}
  \begin{center}
    \includegraphics[width=15cm]{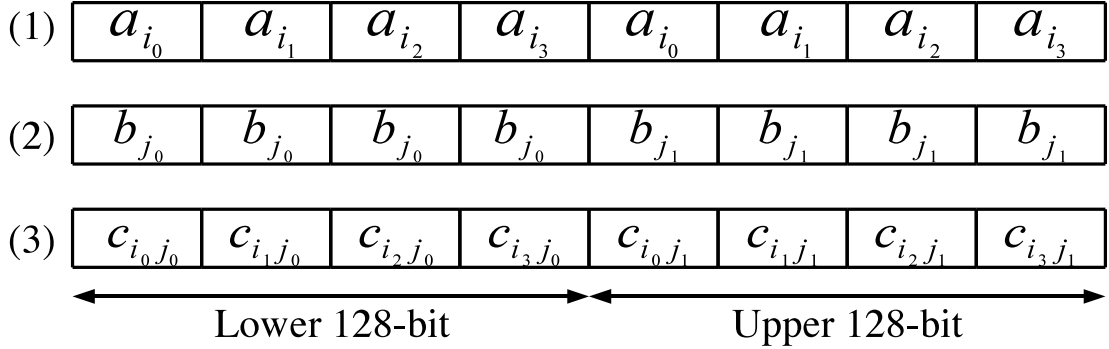}
  \end{center}
  \caption{Data assignments in YMM registers for SIMD
    calculations. The upper panel (1) shows the data assignment of
    four $i$-particles (with indices of $i_0$, $i_1$, $i_2$, and
    $i_3$), where the data are redundantly stored in the lower and
    upper 128-bit in the same order.  The data of two $j$-particles
    (with indices of $j_0$ and $j_1$) are stored in the lower and
    upper 128-bit, respectively, as shown in the middle panel (2). The
    8 values obtained by the operations between the data of four
    $i$-particles and two $j$-particles are stored in the order shown
    in the lower panel (3): $c_{ij} = f(a_i,b_j)$. For example,
    $c_{i_0 j_0}$ is a result of operations between $a_{i_0}$ and
    $b_{j_0}$.}
    \label{fig:assignment}
\end{figure*}

\subsection{Structures for the particle data}

In computing the forces on four $i$-particles from two $j$-particles,
we assign the data of $i$- and $j$-particles to YMM registers in the
way shown in Figure~\ref{fig:assignment}. Suppose that $a$ and $b$ in
Figure \ref{fig:assignment} are $x$-components of $i$- and
$j$-particles, respectively. Subtracting data in the YMM register (1)
of Figure~\ref{fig:assignment} from data in the YMM register (2) of
Figure~\ref{fig:assignment}, we simultaneously obtain $x$-components
of eight relative positions $c$ in the YMM register (3) of
Figure~\ref{fig:assignment}.

In order to effectively realize such SIMD computations with the AVX
instructions, we define the structures for $i$-particles,
$j$-particles and the resulting forces and potentials shown in
List~\ref{list:structures}. Before computing the forces on
$i$-particles, the positions and softening lengths of $i$-particles
are stored in the structure \verb|Ipdata|, and the positions and
masses of $j$-particles are in the structure \verb|Jpdata|. The
resulting forces are stored in the structure \verb|Fodata|. Note that
the structures \verb|Ipdata| and \verb|Fodata| contain the data of
four $i$-particles, while the structure \verb|Jpdata| has the data for
a single $j$-particle.

Note that the positions, softening lengths, and forces of
$i$-particles in the structures \verb|Ipdata| and \verb|Fodata| are
declared as arrays of four single-precision floating-point
numbers. Thus, the data on each array can be suitably loaded onto, or
stored from the lower 128-bit of one YMM register.  The assignment of
the $i$-particles data shown in (1) of Figure~\ref{fig:assignment} can
be realized by loading the data of four $i$-particles onto the lower
128-bit of one YMM register, and copying the data to its upper
128-bit.

As for $j$-particles, since the structure \verb|Jpdata| consists of
four single-precision floating-point numbers, we can load the
positions and the masses of two $j$-particles in one YMM-register at
one time if they are aligned on the 32-byte boundaries. By
broadcasting the $n$-th element ($n=0,1,2$ and 3) in each of the lower
and upper 128-bit to all the other elements, we can realize the
assignment of the $j$-particle data as depicted in (2) of
Figure~\ref{fig:assignment}.

After executing the gravitational force loop over $j$-particles, the
partial forces on four $i$-particles exerted by different sets of
$j$-particles are stored in the upper and lower 128-bit of a YMM
register. Operating sum reduction on the upper and lower 128-bit of
the YMM register, and storing the results into its lower 128-bit, we
can smoothly store the results into the structure \verb|Fodata|.

\begin{lstlisting}[caption={Structures for $i$-particles, $j$-particles, and the resulting forces.}, label={list:structures}]
// structure for i-particles
typedef struct ipdata{
  float x[4];
  float y[4];
  float z[4];
  float eps2[4];
} Ipdata, *pIpdata;

// structure for j-particles
typedef struct jpdata{
  float x, y, z, m;
} Jpdata, *pJpdata;

// structure for the resulting forces
// and potentials of i-particles
typedef struct fodata{
  float ax[4];
  float ay[4];
  float az[4];
  float phi[4];
} Fodata, *pFodata;

\end{lstlisting}

\subsection{Macros for inline assembly codes}

For the readability of the source codes shown below, let us introduce
some preprocessor macros which are expanded into inline assembly
codes. The definitions of the macros used in this paper are given in
List~\ref{list:macros}. For macros with two and three operands, the
results are stored in the second and third one, respectively, and the
other operands are source operands. In these macros, operands named
\verb|src|, \verb|src1|, \verb|src2|, and \verb|dst| designate the
data in XMM or YMM registers, and those named \verb|mem|,
\verb|mem64|, \verb|mem128|, and \verb|mem256| are data in the main
memory or the cache memory, where numbers after \verb|mem| indicate
their size and alignment in bits. Brief descriptions of these macros
are summarized in Table~\ref{tab:macros}.  More detailed explanation
of the AVX instructions can be found in Intel's
website\footnote{\ttfamily http://software.intel.com/en-us/avx/}.

\begin{lstlisting}[caption={Preprocessor macros for inline assembly codes.}, label={list:macros}]
#define VZEROALL asm("vzeroall");
#define VLOADPS(mem256, dst) \
  asm("vmovaps %0, %"dst::"m"(mem256));
#define VSTORPS(reg, mem256) \
  asm("vmovaps %"reg ", %0" ::"m"(mem256));
#define VLOADPS(mem128, dst) \
  asm("vmovaps %0, %"dst::"m"(mem128));
#define VSTORPS(reg, mem128) \
  asm("vmovaps %"reg ", %0" ::"m"(mem128));
#define VLOADLPS(mem64, dst) \
  asm("vmovlps %0, %"dst ", %"dst::"m"(mem64));
#define VLOADHPS(mem64, dst) \
  asm("vmovhps %0, %"dst ", %"dst::"m"(mem64));
#define VBCASTL128(src, dst) \
  asm("vperm2f128 %0, %"src ", %"src \
  ", %"dst " "::"g"(0x00));
#define VCOPYU128TOL128(src,dst) \
  asm("vextractf128 %0, %"src ", %"dst \
  " "::"g"(0x01));
#define VGATHERL128(src1,src2,dst) \
  asm("vperm2f128 %0, %"src2 ", %"src1 \
  ", %"dst " "::"g"(0x02));
#define VCOPYALL(src,dst) \
  asm("vmovaps %0, %"src ", %"dst);
#define VBCAST0(src, dst) \
  asm("vshufps %0, %"src ", %"src \
  ", %"dst " "::"g"(0x00));
#define VBCAST1(src, dst) \
  asm("vshufps %0, %"src ", %"src \
  ", %"dst " "::"g"(0x55));
#define VBCAST2(src, dst) \
  asm("vshufps %0, %"src ", %"src \
  ", %"dst " "::"g"(0xaa));
#define VBCAST3(src, dst) \
  asm("vshufps %0, %"src ", %"src \
  ", %"dst " "::"g"(0xff));
#define VMIX0(src1,src2,dst) \
  asm("vshufps %0, %"src2 ", %"src1 \
  ", %"dst " "::"g"(0x88));
#define VMIX1(src1,src2,dst) \
  asm("vshufps %0, %"src2 ", %"src1 \
  ", %"dst " "::"g"(0xdd));
#define VADDPS(src1, src2, dst) \
  asm("vaddps " src1 "," src2 "," dst);
#define VSUBPS(src1, src2, dst) \
  asm("vsubps " src1 "," src2 "," dst);
#define VSUBPS_M(mem256, src, dst) \
  asm("vsubps %0, %"src ", %"dst \
  " "::"m"(mem256));
#define VMULPS(src1, src2, dst) \
  asm("vmulps " src1 "," src2 "," dst);
#define VRSQRTPS(src, dst) \
  asm("vrsqrtps " src "," dst);
#define VMINPS(src1, src2, dst) \
  asm("vminps "  src1 ", " src2 "," dst);
#define VPSRLD(imm, src1, src2) \
  asm("vpsrld %0, %"src1 ", %"src2::"I"(imm));
#define VPSLLD(imm, src1, src2) \
  asm("vpslld %0, %"src1 ", %"src2::"I"(imm));
#define PREFETCH(mem) \
  asm("prefetcht0 %0"::"m"(mem));
\end{lstlisting}

\begin{table*}[htbp]
  \caption{Descriptions of the macros for inline assembly codes. One `value' denotes a single-precision floating-point number.}
  \label{tab:macros}
  \begin{center}
  \begin{tabular}{|l|l|}
    \hline
    \verb|VZEROALL| & zero out all the YMM registers. \\
    \hline
    \verb|VLOADPS(mem256,dst)| & load eight packed values in
    \verb|mem256| to \verb|dst|. \\    
    \hline
    \verb|VSTORPS(src,mem256)| & store eight packed values in 
    \verb|src| to \verb|mem256|. \\
    \hline
    \verb|VLOADPS(mem128,dst)| & load four packed values in
    \verb|mem128| to \verb|dst|. \\    
    \hline
    \verb|VSTORPS(src,mem128)| & store four packed values in 
    \verb|src| to \verb|mem128|. \\
    \hline
    \verb|VLOADLPS(mem64,dst)| & load two packed values in \verb|mem64|
    to the lower 64-bit of the lower 128-bit in \verb|dst|. \\
    \hline
    \verb|VLOADHPS(mem64,dst)| & load two packed values in \verb|mem64|
    to the upper 64-bit of the lower 128-bit in \verb|dst|. \\
    \hline
    \verb|VBCASTL128(src,dst)| & broadcast data in the lower 128-bit
    of \verb|src| to the lower and upper 128-bit of \verb|dst|. \\
    \hline
    \verb|VCOPYU128TOL128(src,dst)| & copy the upper 128-bit in
    \verb|src| to the lower 128-bit in \verb|dst|. \\
    \hline
    \verb|VGATHERL128(src1,src2,dst)| & copy the lower 128-bit in
    \verb|src1| and \verb|src2| to the upper 128-bit and lower 
    128-bit in \verb|dst|, respectively. \\
    \hline
    \verb|VCOPYALL(src,dst)| & copy 256-bit data from \verb|src| to \verb|dst|. \\
    \hline
    \verb|VBCASTn(src,dst)| & broadcast the n-th element of
    each of the lower and upper 128-bit to all the other elements. \\
    \hline
    \verb|VMIX0(src1,src2,dst)| & operate data as shown in Figure \ref{fig:mix}. \\
    \hline
    \verb|VMIX1(src1,src2,dst)| & operate data as shown in Figure \ref{fig:mix}. \\
    \hline
    \verb|VADDPS(src1,src2,dst)| & add \verb|src1| to \verb|src2|, and
    store the result to \verb|dst|. \\
    \hline
    \verb|VSUBPS(src1,src2,dst)| & subtract \verb|src1| from \verb|src2|,
    and store the result to \verb|dst|. \\
    \hline
    \verb|VSUBPS_M(mem256, src, dst)| & subtract \verb|mem256| from \verb|src|,
    and store the result to \verb|dst|. \\
    \hline
    \verb|VMULPS(src1,src2,dst)| & multiply \verb|src1| by \verb|src2|,
    and store the result to \verb|dst|. \\
    \hline
    \verb|VRSQRTPS(src,dst)| & compute the inverse-square-root of \verb|src|,
    and store the result to \verb|dst|. \\
    \hline
    \verb|VMINPS(src1,src2,dst)| & compare the values in each pair of elements in \verb|src1| and \verb|src2|, and store the not larger ones to \verb|dst|. \\
    \hline
    \verb|VPSRLD(imm,src,dst)| & shift each element in the lower 128-bit of \verb|src| left by \verb|imm| bit, and store the result to \verb|dst|. \\
    \hline
    \verb|VPSRRD(imm,src,dst)| & shift each element in the lower 128-bit of \verb|src| right by \verb|imm| bits, and set the result to \verb|dst|. \\
    \hline
    \verb|PREFETCH(mem)| & prefetch data on \verb|mem| to the cache memory. \\
    \hline
  \end{tabular}
  \end{center}
\end{table*}

\begin{figure}
  \begin{center}
    \includegraphics[width=7.5cm]{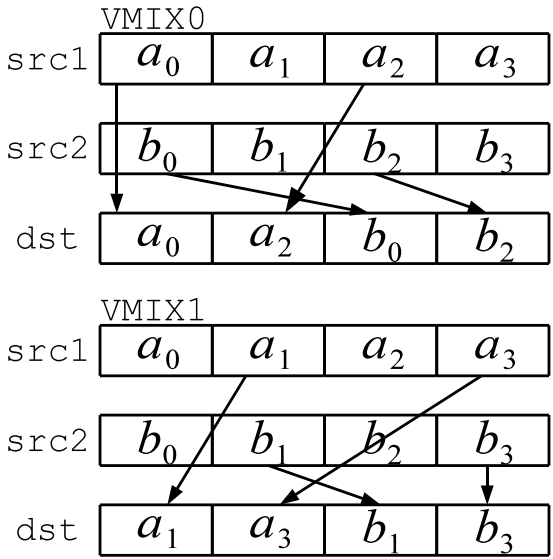}
  \end{center}
  \caption{Instructions {\tt MIX0} and {\tt MIX1}. Each set of four
    boxes indicates the lower (or upper) 128-bit of a YMM
    register. Each box contains a single-precision floating-point
    number.}
  \label{fig:mix}
\end{figure}

Furthermore, we define aliases of XMM and YMM
registers. Table~\ref{tab:alias1} and \ref{tab:alias2} show the
aliases of YMM registers in calculating Newton's force and an
arbitrary shaped central force, respectively. Aliases with suffix
``\verb|_X|'' indicate the lower 128-bit of the original YMM register
which can be used as XMM registers for the SSE instructions. Note that
some of aliases are reused for data other than described in
Table~\ref{tab:alias1} and \ref{tab:alias2}.

\begin{table}[htbp]
  \caption{Aliases of YMM registers for calculating Newton's forces in
    List~\ref{list:newtonforce}.}
  \label{tab:alias1}
  \begin{center}
    \begin{tabular}{|c|c|c|}
      \hline
      Alias&ID& Description \\
      \hline
      \verb|XI|    & \verb|%ymm0|  & \\
      \verb|YI|    & \verb|%ymm1|  &  $x$, $y$, and $z$-coordinates of $i$-particles \\
      \verb|ZI|    & \verb|%ymm2|  & ($x_i$, $y_i$, and $z_i$) \\
      \hline
      \verb|EPS2|  & \verb|%ymm3|  & square of the gravitational softening length ($\epsilon^2$)\\
      \hline
      \verb|AX|    & \verb|%ymm4|  & \\
      \verb|AY|    & \verb|%ymm5|  & forces of $i$-particles\\
      \verb|AZ|    & \verb|%ymm6|  & ($a_{x,i}$, $a_{y,i}$, and $a_{z,i}$) \\
      \hline
      \verb|PHI|   & \verb|%ymm7|  & gravitational potentials of $i$-particles ($\phi_i$) \\
      \hline
      \verb|XJ|    & \verb|%ymm8|  & \\
      \verb|YJ|    & \verb|%ymm9|  & $x$, $y$, and $z$-coordinates of $j$-particles \\
      \verb|ZJ|    & \verb|%ymm10| & ($x_j$, $y_j$, and $z_j$) \\
      \hline
      \verb|MJ|    & \verb|%ymm11| & masses of $j$-particles ($m_j$) \\
      \hline
      \verb|DX|    & \verb|%ymm12| & \\
      \verb|DY|    & \verb|%ymm13| & relative coordinates between $i$- and $j$-particles \\
      \verb|DZ|    & \verb|%ymm14| & ($x_{ij}$, $y_{ij}$, and $z_{ij}$) \\
      \hline
    \end{tabular}
  \end{center}
\end{table}

\begin{table}[htbp]
  \caption{Aliases of YMM registers for calculating an arbitrary force
    shape in List~\ref{list:arbitraryforce}.}
  \label{tab:alias2}
  \begin{tabular}{|c|c|c|}
    \hline
    Alias & ID & Description \\
    \hline
    \verb|X2|    & \verb|%ymm0|  & \\
    \verb|Y2|    & \verb|%ymm1|  & squared inter-particle distances \\
    \verb|Z2|    & \verb|%ymm2|  & \\
    \hline
    \verb|TWO|   & \verb|%ymm3|  & constant value of \verb|2.0| in single-precision \\
    \hline
    \verb|AX|    & \verb|%ymm4|  & \\
    \verb|AY|    & \verb|%ymm5|  & forces of $i$-particles \\
    \verb|AZ|    & \verb|%ymm6|  & \\
    \hline
    \verb|R2CUT| & \verb|%ymm7|  & cutoff radius squared \\
    \hline
    \verb|BUF0|  & \verb|%ymm8|  & \\
    \verb|BUF1|  & \verb|%ymm9|  & buffers used to refer a look-up table \\
    \verb|BUF2|  & \verb|%ymm10| & \\
    \hline
    \verb|MJ|    & \verb|%ymm11| & masses of $j$-particles \\
    \hline
    \verb|DX|    & \verb|%ymm12| & \\
    \verb|DY|    & \verb|%ymm13| & relative coordinates between $i$- and $j$-particles\\
    \verb|DZ|    & \verb|%ymm14| & \\
    \hline
    \verb|ZI|    & \verb|%ymm15| & $z$-components of positions of $i$-particles \\
    \hline
  \end{tabular}
\end{table}

\subsection{Newton's force}
\label{sec:methodnewton}

\begin{figure*}[htbp]
  \begin{center}
    \includegraphics[width=16cm]{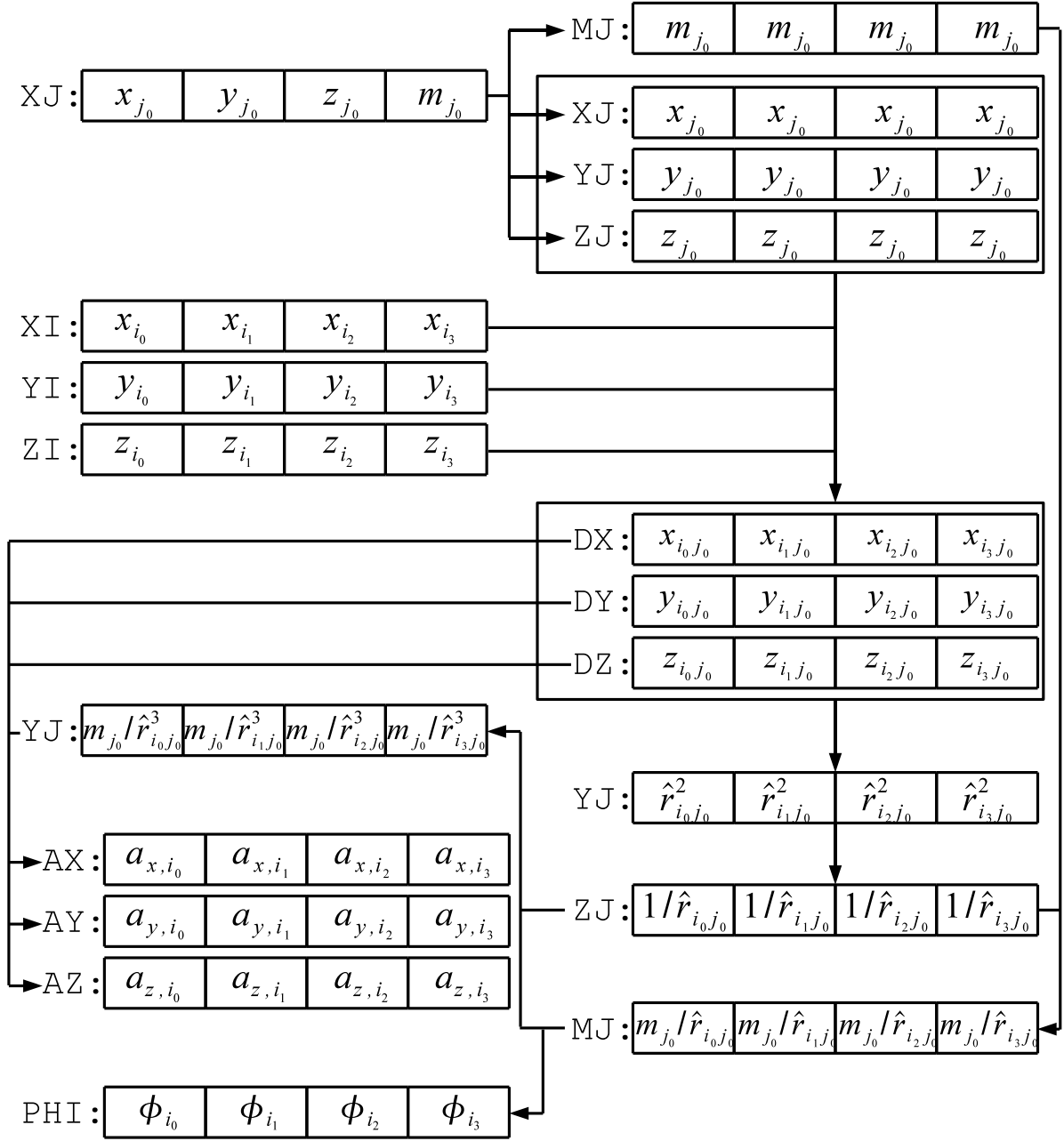}
  \end{center}
  \caption{A schematic illustration of the force loop. Each set of
    four boxes indicates the lower 128-bit of a YMM register. Each box
    contains a single-precision floating-point number. Note that some
    of aliases are reused to store data other than described in
    Table~\ref{tab:alias2}.}
  \label{fig:newtonforce}
\end{figure*}

Figure~\ref{fig:newtonforce} is a schematic illustration of a force
loop to compute the Newton's force on four $i$-particles with AVX
instructions. In this figure, we depict only the lower 128-bit of YMM
registers just for simplicity, while, in actual computation, the upper
128-bit is used to compute forces on the same four $i$-particle
exerted by another $j$-particle.

The overall procedures to calculate the force on four $i$-particles
using AVX instructions are summarized as follows:
\begin{enumerate}
\item [0.] Zero out all the YMM registers, and load the $x$, $y$, and
  $z$ coordinates of four $i$-particles, and squared softening lengths
  to the lower 128-bit of \verb|XI|, \verb|YI|, \verb|ZI|, and
  \verb|EPS2| (i.e. \verb|XI_X|, \verb|YI_X|, \verb|ZI_X|, and
  \verb|EPS2_X|), and copy them to the upper 128-bit of \verb|XI|,
  \verb|YI|, \verb|ZI|, and \verb|EPS2|, respectively.
\item [1.] Load the $x$, $y$, and $z$ coordinates and the masses of
  two $j$-particles to \verb|XJ|.
\item [2.] Broadcast the $x$, $y$, and $z$ coordinates and the masses
  of two $j$-particles in \verb|XJ| to \verb|XJ|, \verb|YJ|,
  \verb|ZJ|, and \verb|MJ|, respectively.
\item [3.] Subtract \verb|XI|, \verb|YI|, and \verb|ZI| from
  \verb|XJ|, \verb|YJ|, and \verb|ZJ| respectively. The results
  ($x_{ij}$, $y_{ij}$, and $z_{ij}$) are stored in \verb|DX|,
  \verb|DY|, and \verb|DZ|, respectively.
\item [4.] Square $x_{ij}$ in \verb|DX|, $y_{ij}$ in \verb|DY|, and
  $z_{ij}$ in \verb|DZ| and sum them up to compute the squared
  distance between two $j$-particles and four $i$-particles. The
  results are stored in the alias \verb|YJ|. The squared softening
  lengths \verb|EPS2| are also added. Eventually, the softened squared
  distances ${\hat{r}_{ij}}^2 \equiv r_{ij}^2+\epsilon^2$ between two
  $j$-particles and four $i$-particles are stored in \verb|YJ|.
\item [5.] Calculate inverse-square-root for $\hat{r}_{ij}^2$ in
  \verb|YJ|, and store the result $1/\hat{r}_{ij}$ in the alias
  \verb|ZJ|.
\item [6.] Multiply $1/\hat{r}_{ij}$ in \verb|ZJ| by $m_j$ in
  \verb|MJ| to obtain $m_j/\hat{r}_{ij}$, and store the results in
  \verb|MJ|.
\item [7.] Accumulate $m_j/\hat{r}_{ij}$ in \verb|MJ| into $\phi_i$ in
  \verb|PHI|.
\item [8.] Square $1/\hat{r}_{ij}$ in \verb|ZJ|, multiply the result
  $1/\hat{r}_{ij}^2$ by $m_j/\hat{r}_{ij}$ in \verb|MJ|, and store
  them $m_j/\hat{r}_{ij}^3$ in \verb|YJ|.
\item [9.] Multiply $x_{ij}$ in \verb|DX|, $y_{ij}$ in \verb|DY|, and
  $z_{ij}$ in \verb|DZ| by $m_j/\hat{r}_{ij}^3$ in \verb|YJ| obtaining
  the forces ($m_jx_{ij}/\hat{r}_{ij}^3$, $m_jy_{ij}/\hat{r}_{ij}^3$,
  and $m_jz_{ij}/\hat{r}_{ij}^3$), and accumulate them into \verb|AX|,
  \verb|AY|, and \verb|AZ|, respectively.
\item [10.] Return to step 1 until all the $j$-particles are
  processed.
\item [11.] Operate sum reduction of partial forces and potentials in
  the lower and upper 128-bits of \verb|AX|, \verb|AY|, \verb|AZ|, and
  \verb|PHI|, and store the results in the lower 128-bit of \verb|AX|,
  \verb|AY|, \verb|AZ|, and \verb|PHI|, respectively.
\item [12.] Store forces and potentials in the lower 128-bit of
  \verb|AX|, \verb|AY|, \verb|AZ|, and \verb|PHI| to the structure
  \verb|Fodata|.
\end{enumerate}

The function \verb|GravityKernel| to compute the forces is shown in
List~\ref{list:newtonforce}. The order of instructions in
List~\ref{list:newtonforce} is slightly different from that described
above in order to obtain high issue rate of the AVX instructions by
optimizing the order of operations so that operands in adjacent
instruction calls do not have dependencies as much as
possible. 
Further optimization is given by explicitly unrolling the force loop,
which does not appear in the list.

\begin{lstlisting}[caption={A force loop to calculate the Newton's force using the AVX instructions.}, label={list:newtonforce}]
void GravityKernel(pIpdata ipdata, pFodata fodata,
                   pJpdata jpdata, int nj)
{
  int j;

  PREFETCH(jpdata[0]);

  VZEROALL;

  VLOADPS(*ipdata->x, XI_X);
  VLOADPS(*ipdata->y, YI_X);
  VLOADPS(*ipdata->z, ZI_X);
  VLOADPS(*ipdata->eps2, EPS2_X);
  VBCASTL128(XI, XI);
  VBCASTL128(YI, YI);
  VBCASTL128(ZI, ZI);
  VBCASTL128(EPS2, EPS2);

  VLOADPS(*(jpdata), XJ);
  jpdata += 2;

  VBCAST1(XJ, YJ);
  VBCAST2(XJ, ZJ);
  VBCAST3(XJ, MJ);
  VBCAST0(XJ, XJ);

  for(j = 0 ; j < nj; j += 2) {
    VSUBPS(YI, YJ, DY);
    VSUBPS(ZI, ZJ, DZ);
    VSUBPS(XI, XJ, DX);

    VMULPS(DZ, DZ, ZJ);
    VMULPS(DX, DX, XJ);
    VMULPS(DY, DY, YJ);

    VADDPS(XJ, ZJ, ZJ);
    VADDPS(EPS2, YJ, YJ);
    VADDPS(YJ, ZJ, YJ);

    VLOADPS(*(jpdata), XJ);
    jpdata += 2;

    VRSQRTPS(YJ, ZJ);

    VMULPS(ZJ, MJ, MJ);
    VMULPS(ZJ, ZJ, YJ);

    VMULPS(MJ, YJ, YJ);
    VSUBPS(MJ, PHI, PHI);

    VMULPS(YJ, DX, DX);
    VMULPS(YJ, DY, DY);
    VMULPS(YJ, DZ, DZ);

    VBCAST1(XJ, YJ);
    VBCAST2(XJ, ZJ);
    VBCAST3(XJ, MJ);
    VBCAST0(XJ, XJ);

    VADDPS(DX, AX, AX);
    VADDPS(DY, AY, AY);
    VADDPS(DZ, AZ, AZ);
  }

  VCOPYU128TOL128(AX, DX_X);
  VADDPS(AX, DX, AX);
  VCOPYU128TOL128(AY, DY_X);
  VADDPS(AY, DY, AY);
  VCOPYU128TOL128(AZ, DZ_X);
  VADDPS(AZ, DZ, AZ);
  VCOPYU128TOL128(PHI, MJ_X);
  VADDPS(PHI, MJ, PHI);

  VSTORPS(AX_X,  *fodata->ax);
  VSTORPS(AY_X,  *fodata->ay);
  VSTORPS(AZ_X,  *fodata->az);
  VSTORPS(PHI_X, *fodata->phi);
}
\end{lstlisting}

\subsection{Central force with an arbitrary shape}
\label{sec:methodarbitrary}

In this section, we describe how to accelerate the computation of
arbitrarily shaped forces $f(r)/r$, using the AVX instructions, where
$f(r)$ is a user-specified function in equation
(\ref{eq:arbitrary_force}).  Note that the inter-particle softening is
also expressed in the force shape function $f(r)$, as well as the long
range cut-off.  Arbitrary shaped softening including the Plummer
softening, $S2$ softening, etc. can be set.  The function $f(r)$ is
assumed to shape; almost constant at $r<\epsilon$, rapidly decreases
at larger $r$, and reaches zero at $r=r_{\rm cut}$.  Such assumptions
are satisfied in the inter-particle force calculations of PPPM or
TreePM methods.

In order to calculate central forces with an arbitrary shape in
equation (\ref{eq:arbitrary_force}), we refer to a pre-calculated
look-up table of $f(r)/r$ and use the linear interpolation between the
sampling points. In \S~\ref{sec:maketable} and \ref{sec:usetable}, we
describe our scheme to construct the look-up table, and procedure to
calculate the force by using the look-up table with the AVX
instructions, respectively.

\subsubsection{Construction of an optimized look-up table}
\label{sec:maketable}

In terms of numerical accuracy, the look-up table is preferred to have
a large number of sampling points between $0\le r\le r_{\rm cut}$. On
the other hand, the size of the look-up table should be as small as
possible to avoid cache misses for fast calculations. Thus, it is
important how to choose sampling points of the look-up table in order
to satisfy such exclusive requirements: accuracy and fast calculations
of forces.

\begin{figure}[htbp]
\leavevmode
\begin{center}
  \includegraphics[scale=1.8]{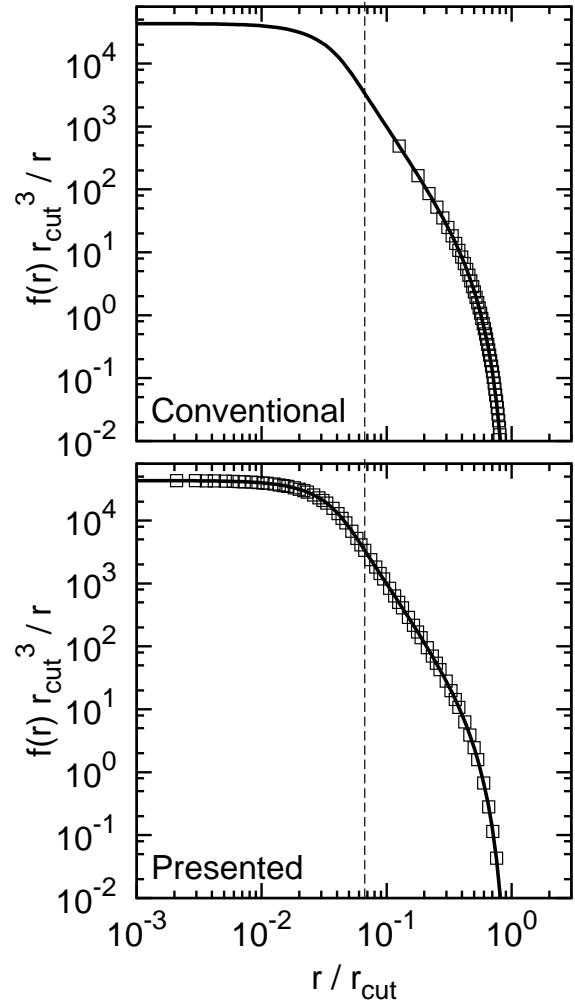}
\end{center}
  \caption{Binning of $f(r)/r$ in the conventional scheme with 64
    constant intervals in $r^2$ (top panel) and in our scheme with
    $E=4$ and $F=2$ (bottom panel) between $[0,r_{\rm cut}]$. Although
    we adopt $F=5$ elsewhere in this paper, we set $F=2$ here for
    viewability. $R(r,\epsilon)-R(r,r_{\rm cut})$ is assumed as a
    functional form of $f(r)$, in which $R(r,\eta)$ is the
    $S2$-profile \citep{Hockney81} (see equation (\ref{eq:s2})). Solid
    lines indicate the shape of $f(r)/r$. Vertical dashed lines in
    both panels are the locations of the gravitational softening
    length $\epsilon$.}
  \label{fig:binning}
\end{figure}

In many previous implementations, sampling points of the look-up table
are chosen so that the sampling points have equal intervals in a
squared inter-particle distance $r^2$ at $0<r<r_{\rm cut}$. However,
the sampling with equal intervals in $r^2$ is not a good choice,
because it has coarser intervals at a smaller inter-particle distance,
and the force shape at $r\lesssim\epsilon$ is poorly sampled if the
number of sampling points is not large enough, while the shape at
$r\simeq r_{\rm cut}$ is sampled fairly well, or even redundantly (see
the top panel of Figure~\ref{fig:binning}). Typically speaking, tens
of thousand sampling points in the region $0<r<r_{\rm cut}$ are
required to assure the sufficient force accuracy if sampling with
equal intervals in $r^2$ is adopted. Such look-up tables need several
hundred kilobytes in single-precision, and do not fit into
a low-level cache memory.

The desirable sampling of the force shapes, therefore, should have
almost equal intervals in $r$ at short distances $r \lesssim
\epsilon$, and intervals proportional to $r$ (or equal intervals in
$\ln r$) at long distances. In the following, we realize such a
sampling by adopting rather a new binning scheme, with which we can
compute the force efficiently.

Here, we consider to construct a look-up table of $f(r)/r$ in the
range of $0<r<r_{\rm cut}$. In our binning scheme, the indices of the
look-up table are calculated by directly extracting the fraction and
the exponent bits of the IEEE754 format of squared inter-particle
distances. First, the squared distance $r^2$ is affine-transformed to
a single-precision floating-point number $s\equiv r^2 (s_{\rm
  max}-2)/r^2_{\rm cut}+2$ so that $s$ is in the range of $s_{\rm min}
< s < s_{\rm max}$, where $s_{\rm min}\equiv 2$ and $s_{\rm max}\equiv
2^{2^{E}}(2-1/2^F)$. Here, $E$ and $F$ are the pre-defined positive
integers, and the numbers of exponent and fraction bits extracted in
computing the indices of the look-up table, respectively.  Binary
expressions of $s_{\rm min}$ and $s_{\rm max}$ in the IEEE754 format
of single-precision (32-bit) in the case of $E=4$ and $F=6$ are shown
in Table~\ref{tab:IEEE754}.  Except that the most significant bit of
the exponent part is always 1, all the bits of $s_{\rm min}$ are 0,
and as for $s_{\rm max}$, only the lower $E$ bits of the exponent and
the higher $F$ bits of the fraction are 1. Next, the indices of the
look-up table for the squared distances $r^2$ or $s$ are computed by
extracting the lower $E$ bits of the exponent and the higher $F$ bits
of the fraction of $s$ (underlined portion of exponent and fraction
bits in Table~\ref{tab:IEEE754}) and reinterpreting it as an
integer. This procedure can be done by applying a logical right shift
by $23-F$ bits, and a bitwise-logical AND with $2^{E+F}-1$ to $s$. It
should be noted that the resulting size of the look-up table is
$2^{E+F}$.

\bigskip

\begin{table}[htbp]
  \begin{center}
    \caption{$s$-values, their exponent and fraction bits in the
      IEEE754 expressions, and their indices in the table for $r=0$,
      $r_{\rm cut}/2$ and $r_{\rm cut}$ in the case of $E=4$ and $F=6$
      (underlined portion of exponent and fraction bits). }
    \label{tab:IEEE754}
  \begin{tabular}{c||c|c|c|c}
    \hline
    $r$ & $s$ & exponent bits & fraction bits & index\\
    \hline
    0&$2$ ($s_{\rm min}$) & 1000\underline{0000} & \underline{000000}00000000000000000 & 0\\ 
    \hline
    $r_{\rm cut}/2$ & $3.2514\times10^4$ & 1000\underline{1101} & \underline{111111}00000001100000000 & 895 \\
    \hline 
    $r_{\rm cut}$ &
    \begin{tabular}{c}
      $1.3005\times 10^5$ \\ ($s_{\rm max}$) \\
    \end{tabular}
    & 1000\underline{1111} & \underline{111111}00000000000000000 & 1023 \\
    \hline
  \end{tabular}
  \end{center}
\end{table}

\bigskip

An affine-transformed squared distance at a sampling point with an
index specified by a lower $E$ exponent bits $b_{\rm E}$ and an upper
$F$ fraction bits $b_{\rm F}$ is expressed as
\begin{equation}
  s_{b_{\rm E},b_{\rm F}} = 2^{b_{\rm E}+1}\left(1+\frac{b_{\rm
      F}}{2^F}\right) \;\; \left(0 \le b_{\rm E} < 2^E, \; 0 \le b_{\rm
    F} < 2^F \right).
\end{equation}
The ratio between inter-particle distances whose affine-transformed
squared distances are $s_{(b_{\rm E}+1),b_{\rm F}}$ and $s_{b_{\rm
    E},b_{\rm F}}$ is given by
\begin{eqnarray}
  \frac{r_{(b_{\rm E}+1),b_{\rm F}}}{r_{b_{\rm E},b_{\rm F}}} &=&
  \left(\frac{s_{(b_{\rm E}+1),b_{\rm F}}-2}{s_{b_{\rm E},b_{\rm
        F}}-2}\right)^{1/2} \nonumber \\
  &\simeq& 2^{1/2}, \label{eq:comp_e}
\end{eqnarray}
where $b_{\rm E} \gg 1$ is assumed for the last approximation. 
The interval between inter-particle distances whose
affine-transformed distances are $s_{b_{\rm E},(b_{\rm F}+1)}$ and
$s_{b_{\rm E},b_{\rm F}}$ is calculated as
\begin{eqnarray}
  r_{b_{\rm E},(b_{\rm F}+1)} - r_{b_{\rm E},b_{\rm F}} &=&
  \left(\frac{s_{b_{\rm E},(b_{\rm F}+1)}-2}{s_{\rm
      max}-2}\right)^{1/2} - \left(\frac{s_{b_{\rm E},b_{\rm
        F}}-2}{s_{\rm max}-2}\right)^{1/2} \nonumber \\ &\simeq&
  \frac{1}{(2^F+b_{\rm F})^{1/2}} \left(\frac{2^{b_{\rm
        E}+1}/2^{F+2}}{s_{\rm max}-2}\right)^{1/2}, \label{eq:comp_f}
\end{eqnarray}
where we also assume $b_{\rm E} \gg 1$ and $F \gg 1$ for the last
approximation. Therefore, the sampling points with the same fraction
bits are distributed uniformly in logarithmic scale, and those with
the same exponent bits are aligned uniformly in linear scale unless
the fraction bit is small.

\begin{figure}[htbp]
\leavevmode
\begin{center}
  \includegraphics[scale=1.8]{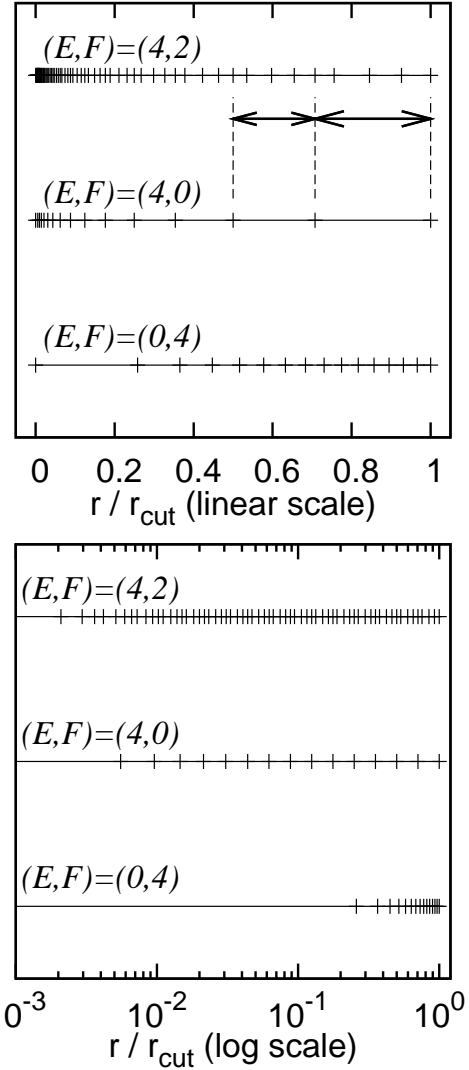}
\end{center}
\caption{Sampling points of an inter-particle distance for a look-up
  table in various cases of pre-defined integers $E$ and $F$. The top
  and bottom panels take horizontal axes in linear and logarithmic
  scales, respectively.}
\label{fig:comp_ef}
\end{figure}

As an example, we illustrate how the sampling points of the look-up
table depend on the pre-defined integers $E$ and $F$ in
Figure~\ref{fig:comp_ef}. We first see the cases in which either of
$E$ and $F$ is zero, in order to see the roles of the integers $E$ and
$F$. As seen in Figure~\ref{fig:comp_ef}, the intervals of sampling
points are roughly uniform in linear scale for the case $E=0$ (the
bottom line in the top panel), and uniform in logarithmic scale for
the case $F=0$ (the middle line in the bottom panel), unless $r/r_{\rm
  cut}$ is small. As expected above, the integers $E$ and $F$ control
the number of sampling points in logarithmic and linear scales,
respectively.

By comparing the sampling points with $(E,F)=(4,0)$ and those with
$(4,2)$ (see the top panel of Figure~\ref{fig:comp_ef}), it can be
seen that all intervals of the sampling points with $(E,F)=(4,0)$
(indicated by the vertical dashed lines and double-headed arrows) are
divided nearly equally into $2^F=4$ regions by the sampling points
with $(E,F)=(4,2)$. Thus, our binning scheme is a hybrid of the linear
and logarithmic binning schemes.

\begin{figure}[htbp]
\leavevmode
\begin{center}
  \includegraphics[scale=1.8]{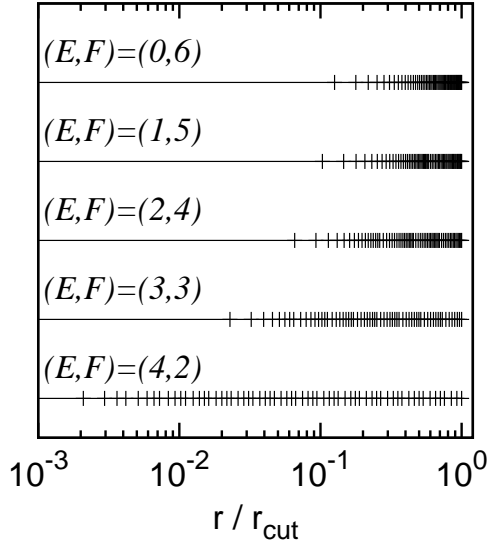}
\end{center}
\caption{Comparison of binning among the same number of sampling
  points in various cases of the integers $E$ and $F$.}
\label{fig:decide_ef}
\end{figure}

Figure~\ref{fig:decide_ef} shows the comparison of the several binning
in which the number of sampling points is fixed to $2^{E+F}=2^6$. One
can see that the binning with $(E,F)=(4,2)$ has sufficient sampling
points in the range of $10^{-3}\le r/r_{\rm cut}\le 10^0$, whereas the
binning with the other sets of $(E,F)$ only samples the region of
$10^{-2}\le r/r_{\rm cut}<10^0$. The number of the extracted exponent
bit $E$ should be large enough so that the scale of the softening
length should be sufficiently resolved. For example, if
$\epsilon/r_{\rm cut}\lesssim 10^{-2}$, $E$ should be set to at least
equal to or larger than $4$.

In List~\ref{list:bit_binning}, we present routines for constructing
the look-up table. In our implementation, the look-up table contains
two values: one is the force at a sampling point $r_k$,
\begin{equation}
  G^0_k = \frac{f(r_k)}{r_k},
\end{equation}
and the other is its difference from the next sampling point $r_{k+1}$
divided by the interval of the affine-transformed squared distance
\begin{equation}
 G^1_k = \frac{G^0_{k+1}-G^0_k}{s_{k+1}-s_k}
\end{equation}
where subscript $k$ indicates indices of the look-up table, and is
expressed as $k=2^{F} \times b_{\rm E} + b_{\rm F}$. Using these two
values, we can compute the linear interpolation of $f(r)/r$ at a
radius $r$ with $r_{k}\le r \le r_{k+1}$ by $G^0_k + (s-s_k)G^1_k$.
The $G^0_k$ and $G^1_k$, are stored in a two-dimensional array
declared as \verb|Force_table[TBL_SIZE][2]|, where \verb|TBL_SIZE| is
the number of the sampling points ($2^{E+F}$) and the values of the
$G^0_k$ and $G^1_k$ are stored in the \verb|Force_table[k][0]| and
\verb|Force_table[k][1]|, respectively.  Since the values of $G^0_k$
and $G^1_k$ are stored in the adjacent memory address, we can avoid
the cache misses in computing the linearly interpolated values of
$f(r)/r$.

\begin{lstlisting}[caption={Implementation of the construction of the look-up table.}, label={list:bit_binning}]
#define EXP_BIT (4)
#define FRC_BIT (6)
#define TBL_SIZE (1 << (EXP_BIT+FRC_BIT)) // 1024

extern float Force_table[TBL_SIZE][2]; // 8 kB

union pack32{
  float f;
  unsinged int u;
};

void generate_force_table(float rcut)
{
  unsigned int tick;
  float fmax, r2scale, r2max;
  union pack32 m32;
  
  float force_func(float);
  
  tick = (1 << (23-FRC_BIT));
  fmax = (1 << (1<<EXP_BIT))*(2.0-1.0/(1<<FRC_BIT));
  r2max = rcut*rcut;
  r2scale = (fmax-2.0f)/r2max;
  
  for(i=0,m32.f=2.0f;i<TBL_SIZE;i++,m32.u+=tick) {
    float f, r2, r;
    
    f=m32.f;
    r2 = (f-2.0)/r2scale;
    float r = sqrtf(r2);
    Force_table[i][0] = force_func(r);
  }
  
  for(i=0,m32.f=2.0f;i<TBL_SIZE-1;i++) {
    float x0 = m32.f;
    m32.u += tick;
    float x1 = m32.f;
    float y0 = Force_table[i][0];
    float y1 = (i==TBL_SIZE-1) ? 0.0
                               : Force_table[i+1][0];
    Force_table[i][1] = (y1-y0)/(x1-x0);
  }
  Force_table[i][1] = 0.0f;
}
\end{lstlisting}

In Figure~\ref{fig:binning}, we compare the conventional binning with
equal intervals in squared distances to our binning with $E=4$ and
$F=2$ (i.e. 64 sampling points), for the $S2$-force shape
\citep{Hockney81} used in the PPPM scheme. Although we adopt $F=5$ in
the rest of this paper, we set $F=2$ here just for good visibility of
the difference of the two binning schemes. It should be noted that the
number of sampling points is the same (64) in both schemes. Compared
with the conventional binning scheme in the top panel, our binning
scheme can faithfully reproduce the given functional form even at
distances smaller than the gravitational softening length.

\subsubsection{Procedure of force calculation}
\label{sec:usetable}

In calculating the arbitrary central forces, the data of $i$- and
$j$-particles are stored in the structures \verb|Ipdata| and
\verb|Jpdata|, respectively, in the same manner as described in the
case for calculating the Newton's force, except that the coordinates
of $i$- and $j$-particles are scaled as
\begin{equation}
  \tilde{{\bm r}}_i = \frac{{\bm r}_i}{r_{\rm cut}/\sqrt{s_{\rm
        max}-2}}, 
  \label{eq:scalingposition}
\end{equation}
so that we can quickly compute the affine-transformed squared
inter-particle distances between $i$- and $j$-particles. As in the
case of the Newton's force, we compute the forces of four
$i$-particles exerted by two $j$-particles using the AVX
instructions. Using the scaled positions of the particles, the
calculation of the forces is performed in the force loop as follows;
\begin{enumerate}
\item Calculate an affine-transformed distance between $i$- and
  $j$-particles, $s$, as
  \begin{equation}
    s = \min \left(|\tilde{\bm r}_j-\tilde{\bm r}_i|^2+2, s_{\rm
      max}\right),
  \end{equation}
  where the function ``$\min$'' returns the minimum value among
  arguments.
\item Derive an index $k$ of the look-up table from the
  affine-transformed squared distance, $s$, computed in the previous
  step by applying a bitwise-logical right shift by $23-F$ bits and
  reinterpreting the result as an integer.
\item Refer to the look-up table to obtain $G^0_k$ and $G^1_k$. Note
  that the address of the pointer to \verb|fcut| is decremented by
  {\tt 1<<(30-(23-F))} in advance (see line 24 in
  List~\ref{list:arbitraryforce}) to correct the effect of the most
  significant exponential bit of $s$.
\item Derive an affine-transformed distance $s_k$ that corresponds to
  the $k$-th sampling point $r_k$ by applying a bitwise-logical left
  shift by $23-F$ bits to $k$ and reinterpreting the result as a
  single-precision floating-point number.
\item Compute the value of $f(|{\bm r}_j-{\bm r}_i|)/|{\bm r}_j-{\bm
  r}_i|$ by the linear interpolation of $G^0_k$ and $G^0_{k+1}$.
  Using the values of $G^0_k$ and $G^1_k$, the interpolation can be
  performed as
  \begin{equation}
    \frac{f(|{\bm r}_j-{\bm r}_i|)}{|{\bm r}_j-{\bm r}_i|} = G^0_k +
    G^1_k \left(s - s_k\right).
  \end{equation}
\item Accumulate scaled ``forces'' on $i$-particles as
  \begin{equation}
    \tilde{\bm a}_i = \sum_j^N m_j \frac{f(|{\bm r}_j-{\bm
        r}_i|)}{|{\bm r}_j-{\bm r}_i|}(\tilde{\bm r}_j-\tilde{\bm
      r}_i)
  \end{equation}
\end{enumerate}

After the force loop, the scaled ``forces'' are rescaled back as 
\begin{equation}
 {\bm a}_i = \frac{r_{\rm cut}}{\sqrt{s_{\rm max}-2}}\tilde{\bm a}_i.
\end{equation}

The actual code of the force loop for the calculation of the central
force with an arbitrary force shape is shown in
List~\ref{list:arbitraryforce}. Note that bitwise-logical shift instructions
such as \verb|VPSRLD| and \verb|VPSLLD| can be operated only to XMM
registers or the lower 128-bit of YMM registers. In order to operate
bitwise-logical shift instructions to data in the upper 128-bit of a YMM
register, we have to copy the data to the lower 128-bit of another YMM
register. Bitwise-logical shift operations to the upper 128-bit of YMM
registers are supposed to be implemented in the future AVX2 instruction
set. Also note that we cannot refer to the look-up table in a SIMD
manner and have to issue the \verb|VLOADLPS| and \verb|VLOADHPS|
instructions one by one (see lines 89--92 and 94--97 in
List~\ref{list:arbitraryforce}). Except for those operations, all the
other calculations are performed in a SIMD manner using the AVX
instructions.

\begin{lstlisting}[caption={Implementation of arbitrary force calculation using AVX instructions.}, label={list:arbitraryforce}]
#define FRC_BIT (6)
#define ALIGN32 __attribute__ ((aligned(32)))
#define ALIGN64 __attribute__ ((aligned(64)))

typedef float  v4sf __attribute__ ((vector_size(16)));
typedef struct ipdata_reg{
  float x[8];
  float y[8];
} Ipdata_reg, *pIpdata_reg;

void GravityKernel(pIpdata ipdata, 
                    pJpdata jp,
                    pFodata fodata,
                    int nj,
                    float fcut[][2],
                    v4sf r2cut, v4sf accscale)
{
  int j;
  unsigned long int ALIGN64 idx[8]
  = {0, 0, 0, 0, 0, 0, 0, 0};
  Ipdata_reg ALIGN32 ipdata_reg;
  static v4sf two = {2.0f, 2.0f, 2.0f, 2.0f};

  fcut -= (1<<(30-(23-FRC_BIT)));

  VZEROALL;

  VLOADPS(ipdata->x[0], X2_X);
  VLOADPS(ipdata->y[0], Y2_X);
  VLOADPS(ipdata->z[0], Z2_X);
  VLOADPS(r2cut, R2CUT_X);
  VLOADPS(two, TWO_X);
  VBCASTL128(X2, X2);
  VSTORPS(X2, ipdata_reg.x[0]);
  VBCASTL128(Y2, Y2);
  VSTORPS(Y2, ipdata_reg.y[0]);
  VBCASTL128(Z2, ZI);
  VBCASTL128(R2CUT, R2CUT);
  VBCASTL128(TWO, TWO);

  VLOADPS(*jp, MJ);
  jp += 2;

  VBCAST0(MJ, X2);
  VBCAST1(MJ, Y2);
  VBCAST2(MJ, Z2);

  VSUBPS_M(*ipdata_reg.x, X2, DX);
  VMULPS(DX, DX, X2);
  VADDPS(TWO, X2, X2);

  VSUBPS_M(*ipdata_reg.y, Y2, DY);
  VMULPS(DY, DY, Y2);
  VADDPS(X2, Y2, Y2);
    
  VSUBPS(ZI, Z2, DZ);
  VMULPS(DZ, DZ, Z2);
  VADDPS(Y2, Z2, Y2);
    
  VBCAST3(MJ, MJ);
  VMULPS(MJ, DX, DX);
  VMULPS(MJ, DY, DY);
  VMULPS(MJ, DZ, DZ);

  VMINPS(R2CUT, Y2, Z2);

  for(j = 0; j < nj; j += 2){
    VLOADPS(*jp, MJ);
    jp += 2;

    VCOPYU128TOL128(Z2, Y2_X);
    VPSRLD(23-FRC_BIT, Y2_X, Y2_X);
    VPSRLD(23-FRC_BIT, Z2_X, X2_X);    
    
    VSTORPS(X2_X, idx[0]);
    VSTORPS(Y2_X, idx[4]);

    VPSLLD(23-FRC_BIT, Y2_X, Y2_X);
    VPSLLD(23-FRC_BIT, X2_X, X2_X);

    VGATHERL128(Y2, X2, Y2);
    VSUBPS(Y2, Z2, Z2);

    VBCAST0(MJ, X2);
    VBCAST1(MJ, Y2);
    VSUBPS_M(*ipdata_reg.x, X2, X2);
    VSUBPS_M(*ipdata_reg.y, Y2, X2);

    VLOADLPS(*fcut[idx[4]], BUF0_X);
    VLOADHPS(*fcut[idx[5]], BUF0_X);
    VLOADLPS(*fcut[idx[0]], BUF1_X);
    VLOADHPS(*fcut[idx[1]], BUF1_X);
    VGATHERL128(BUF0, BUF1, BUF1);
    VLOADLPS(*fcut[idx[6]], BUF2_X);
    VLOADHPS(*fcut[idx[7]], BUF2_X);
    VLOADLPS(*fcut[idx[2]], BUF0_X);
    VLOADHPS(*fcut[idx[3]], BUF0_X);
    VGATHERL128(BUF2, BUF0, BUF2);
    VMIX1(BUF1, BUF2, BUF0);
    VMIX0(BUF1, BUF2, BUF2);

    VMULPS(Z2, BUF0, BUF0);

    VBCAST2(MJ, Z2);
    VBCAST3(MJ, MJ);
    VSUBPS(ZI, Z2, Z2);

    VADDPS(BUF0, BUF2, BUF2);
    VMULPS(BUF2, DX, DX);
    VMULPS(BUF2, DY, DY);
    VMULPS(BUF2, DZ, DZ);

    VADDPS(DX, AX, AX);
    VADDPS(DY, AY, AY);
    VADDPS(DZ, AZ, AZ);

    VCOPYALL(X2, DX);
    VCOPYALL(Y2, DY);
    VCOPYALL(Z2, DZ);

    VMULPS(X2, X2, X2);
    VMULPS(Y2, Y2, Y2);
    VMULPS(Z2, Z2, Z2);

    VADDPS(TWO, X2, X2);
    VADDPS(Z2, Y2, Y2);
    VADDPS(X2, Y2, Y2);
    
    VMULPS(MJ, DX, DX);
    VMULPS(MJ, DY, DY);
    VMULPS(MJ, DZ, DZ);
    VMINPS(R2CUT, Y2, Z2);
  }
  VCOPYU128TOL128(AX, X2_X);
  VADDPS(AX, X2, AX);
  VCOPYU128TOL128(AY, Y2_X);
  VADDPS(AY, Y2, AY);
  VCOPYU128TOL128(AZ, Z2_X);
  VADDPS(AZ, Z2, AZ);

  VMULPS_M(accscale, AX_X, AX_X);
  VMULPS_M(accscale, AY_X, AY_X);
  VMULPS_M(accscale, AZ_X, AZ_X);

  VSTORPS(AX_X, *fodata->ax);
  VSTORPS(AY_X, *fodata->ay);
  VSTORPS(AZ_X, *fodata->az);
}
\end{lstlisting}
Although the AVX instruction set takes the non-destructive
3-operand form, the copy instruction between registers
appeared in the code above, which was intended to avoid
the inter-register dependencies.

\subsection{Parallelization on multi-core processors}

On multi-core processors, we can parallelize the calculations of the
forces of $i$-particles for both of the Newton's force and arbitrary
central forces using the {\tt OpenMP} programming interface by
assigning a different set of four $i$-particles onto each processor
core. List~\ref{list:parallel} shows a code fragment for the
parallelization of the computations of the Newton's force.  The
calculation of an arbitrary force can be parallelized similarly to
that of Newton's force.

\begin{lstlisting}[caption={Code fragment to parallelize the calculations using {\tt OpenMP} programming interface. },label={list:parallel}]
#define ISIMD 4

extern Ipdata ipos[NI_MEMMAX / ISIMD];
extern Jpdata jpos[NJ_MEMMAX];
extern Fodata iacc[NI_MEMMAX / ISIMD];

int nig = ni / ISIMD + (ni % ISIMD ? 1 : 0)
  
#pragma omp parallel for
for(i = 0; i < nig; i++)
  GravityKernel(&ipos[i], &iacc[i], jpos, nj);
\end{lstlisting}

\subsection{Application programming interfaces}

With the implementations of the force calculation accelerated with the
AVX instructions described above, we develop a set of application
programming interfaces (APIs) for $N$-body simulations, which is
compatible to GRAPE-5 library\footnote{\ttfamily
  http://www.kfcr.jp/downloads/g7pkg2.2.1/g5user.pdf}, except that our
library do not support functions to search for neighbours of a given
particle. The APIs are shown in List~\ref{list:api}. \verb|g5_set_xmj|
sends the data of $j$-particles to the array of the structure
\verb|Jpdata|. \verb|g5_calculate_force_on_x| sends the data of
$i$-particles to the array of the structure \verb|Ipdata|, and
computes the forces and potentials of $i$-particles and returns them
into the arrays \verb|ai| and \verb|pi|, respectively.

In the function \verb|g5_open|, we derive statistical bias of the fast
approximation of inverse-square-root, \verb|VRSQRTPS| instruction. As
\cite{Nitadori06} reported, the results of this instruction contains a
bias which is implementation-dependent. We statistically correct this
bias in the same way as \cite{Nitadori06}.

Softening length and the number of $j$-particles are set by the
functions \verb|g5_set_eps_to_all| and \verb|g5_set_n|,
respectively. \verb|g5_close| does nothing and is just for
compatibility with the GRAPE-5 library.

List~\ref{list:sample} shows a code fragment to perform an $N$-body
simulation, using this APIs.

\begin{lstlisting}[caption={APIs.},label={list:api}]
void g5_open(void);
void g5_close(void);
void g5_set_eps_to_all(double eps);
void g5_set_n(int nj);
void g5_set_xmj(int adr,
                int nj,
                double (*xj)[3],
                double *mj);
void g5_calculate_force_on_x(double (*xi)[3],
                             double (*ai)[3],
                             double *pi,
                             int ni);
\end{lstlisting}

\begin{lstlisting}[caption={Sample code.},label={list:sample}]
int n;             // The number of particles
double m[NMAX];    // Mass
double x[NMAX][3]; // Position
double v[NMAX][3]; // Velocity
double a[NMAX][3]; // Force
double p[NMAX];    // Potential
double t;          // Time
double tend;       // Time at the finish time
double dt;         // Timestep
void time_integrator(int,
                     double (*)[3],
                     double (*)[3],
                     double (*)[3]
                     double);
                   // Function for time integration

g5_open();
g5_set_eps_to_all(eps);
g5_set_n(n);
while(t < tend){
  g5_set_xmj(0,n,x,m);
  g5_calculate_force_on_x(x,a,p,n);
  time_integrator(n,x,v,a,dt);
  t += dt;
}
g5_close();
\end{lstlisting}

For the version of arbitrary force shape, we provide a new API call to
set the force-table through a function pointer, which is not compatible
to the GRAPE-5 API.

\section{Accuracy}
\label{sec:accuracy}

\subsection{Newton's force}

We investigate accuracy of forces and potentials obtained by our
implementation for Newton's force. For this purpose, we compute the
forces and potentials of particles in the Plummer models using our
implementations and compare them with those computed fully in
double-precision floating-point numbers without any explicit use of
the AVX instructions. For the calculations of the forces and the
potentials, we adopt the direct particle-particle method and the
softening length of $4r_{\rm v}/N$, where $r_{\rm v}$ is a virial
radius of the Plummer model and $N$ is the number of particles.

\begin{figure*}[htbp]
\leavevmode
\begin{center}
  \includegraphics[width=16cm]{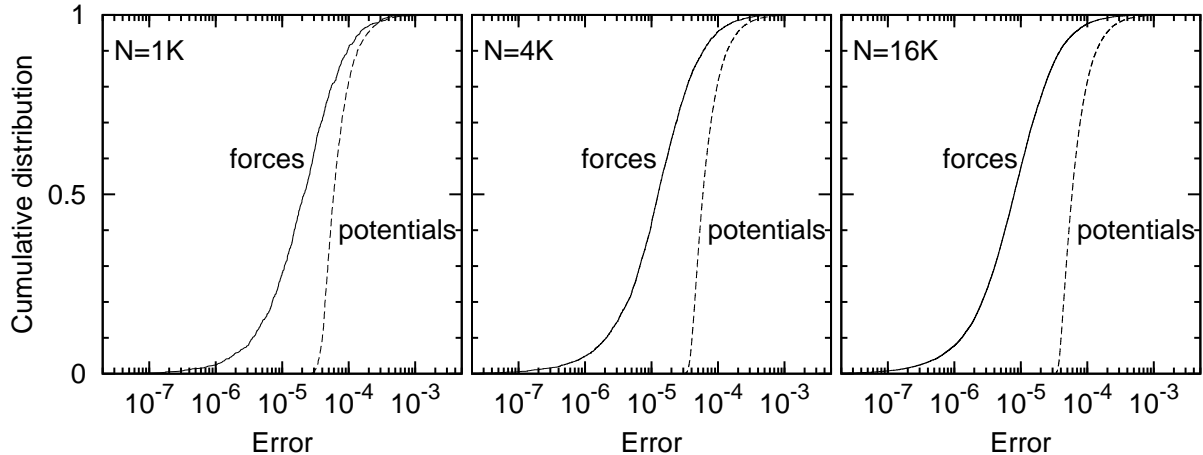}
\end{center}
\caption{Cumulative distribution of errors in forces (solid curves)
  and potentials (dashed curves) of particles in Plummer models with
  $N=1$K, $4$K, and $16$K, where $1$K is equal to $1024$. Softening
  lengths are set to $4r_{\rm v}/N$. The errors are calculated as
  $|{\bm a}_{\rm AVX}-{\bm a}_{\rm DP}|/|{\bm a}_{\rm DP}|$ and
  $|\phi_{\rm AVX}-\phi_{\rm DP}|/|\phi_{\rm DP}|$, where $\bm{a}$ is
  force, $\phi$ potential. The subscripts of ``AVX'' and ``DP''
  indicate the forces and potentials obtained as our implementation in
  section~\ref{sec:methodnewton}, and those obtained by performing all
  the calculations in double-precision, respectively. We deal
  self-interactions as described in the text.}
  \label{fig:newton_error}
\end{figure*}

Figure~\ref{fig:newton_error} shows the cumulative distribution of
relative errors in the forces and the potentials of particles,
\begin{equation}
 \frac{|{\bm a}_{\rm AVX}-{\bm a}_{\rm DP}|}{|{\bm a}_{\rm DP}|},
\end{equation}
and 
\begin{equation}
 \frac{|\phi_{\rm AVX}-\phi_{\rm DP}|}{|\phi_{\rm DP}|},
\end{equation}
where ${\bm a}_{\rm AVX}$ and $\phi_{\rm AVX}$ are the force and the
potential calculated using our implementation, and ${\bm a}_{\rm DP}$
and $\phi_{\rm DP}$ are those computed fully in double-precision. It
can be seen that most of the particles have errors less than
$10^{-4}$. These errors primarily come from the approximate
inverse-square-root instruction \verb|VRSQRTPS|, whose accuracy is
about 12-bit, and consistent with the typical errors of $\simeq
10^{-4}$.

While the errors of the forces are distributed down to less than
$10^{-7}$, the errors of the potentials are mostly larger than $\simeq
3 \times 10^{-5}$.  It can be ascribed to the way of excluding the
contribution of self-interaction to the potentials. In computing a
potential of the $i$-th particle, we accumulate the contribution from
particle pairs between the $i$-th particle and all the particles
including itself, and then subtract the contribution of the potential
between the $i$-th particle and itself, $-m_i/\epsilon$ to finally
obtain the correct potential of the $i$-th particle. Note that the
potential between the $i$-th particle and itself is largest among the
potentials between the $i$-th particle and all the particles, since
the separation between $i$-particle and itself is zero. Thus, the
subtraction of the ``potential'' due to the self-interaction causes
the cancellation of the significant digits, and consequently degrades
the accuracy of the potentials.

A remedy for such degradation of the accuracy is to avoid the
self-interaction in the force loop. In fact, we do so in calculating
the potentials in double-precision ($\phi_{\rm DP}$) in
Figure~\ref{fig:newton_error}. However, such treatment requires
conditional bifurcation inside the force loop, and significantly
reduces the computational performance. The potentials of particles are
usually necessary only for checking the total energy conservation, and
the accuracy obtained in our implementation is sufficient for that
purpose. For these reasons, we choose the original way to compute the
potentials of particles in our implementation.

\subsection{Central force with an arbitrary shape}

In order to see accuracies of central forces with an arbitrary shape
obtained in our implementation, we choose a force shape which is
frequently adopted in cosmological $N$-body simulations using PPPM or
TreePM methods. Such methods are comprised of the particle--mesh (PM)
and the particle--particle (PP) parts which compute long- and
short-range components of inter-particle forces, respectively. Our
implementation of the calculation of arbitrarily-shaped central forces
can accelerate the calculation of the PP part, in which the force
shape is different from the Newton's force and is expressed as
\begin{equation}
  f(r) = R(r,\epsilon) -  R(r,r_{\rm cut}),
  \label{eq:s2pp}
\end{equation}
where $R(r,a)$ is the so-called $S2$-profile with a softening length of
$a$ \citep{Hockney81} given by
\begin{equation}
  \displaystyle
  R(r,a) = \left\{
  \begin{array}{l}
    \bigl(224\xi-224\xi^3+70\xi^4+48\xi^5-21\xi^6\bigr)/35a^2\\
    \;\;\; \mbox{for ($0 \le \xi < 1$)} \\
    \bigl(12/\xi^2-224+896\xi-840\xi^2+224\xi^3+70\xi^4\\
    \;\;\; -48\xi^5+7\xi^6\bigr)/35a^3 \; \mbox{for ($1 \le \xi < 2$)} \\
    \frac{1}{r^2} \; \mbox{for ($2 \le \xi$)} \\
  \end{array}
  \right. . \label{eq:s2}
\end{equation}

We calculate forces exerted between $4$K particle pairs with various
separations uniformly distributed in $\ln(r)$ in the range of $5\times
10^{-3} < r/r_{\rm cut} < 1$ using our implementation described in
section~\ref{sec:methodarbitrary}, where $1$K is equal to $1024$. We
set $\epsilon$ and $r_{\rm cut}$ to $3.125 \times 10^{-3}$ and $4.6875
\times 10^{-2}$, and masses to unity. In creating the look-up table of
the force shape, we set $E=4$ and $F=5$.

\begin{figure}[htbp]
  \begin{center}
    \includegraphics[scale=1.8]{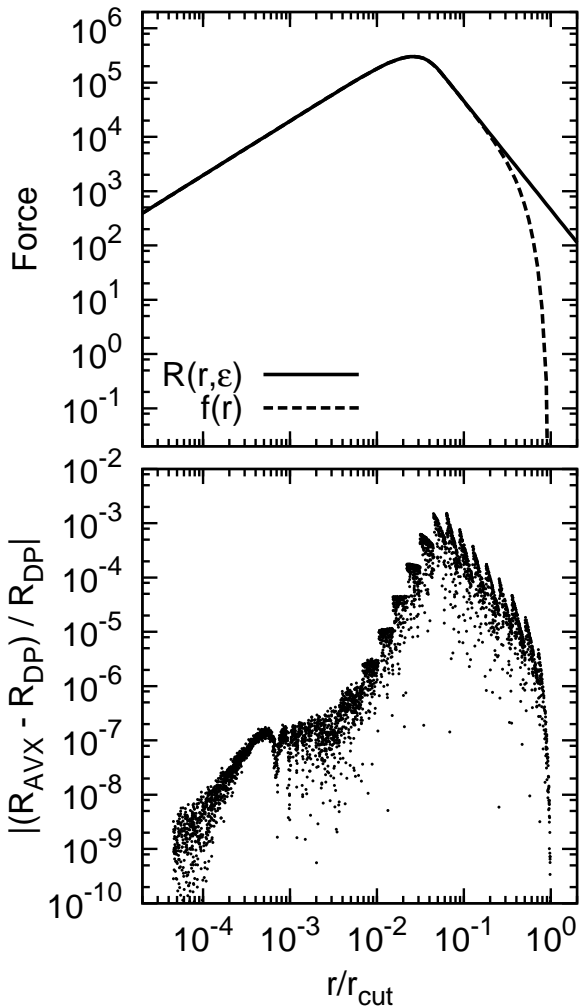}
  \end{center}
  \caption{Shape of $R(r,\epsilon)$ and $f(r)$ (upper panel) and the
    relative errors of forces of particle pairs with a separation $r$
    (bottom panel) as a function of $r/r_{\rm cut}$, where the forces
    include both PP and PM parts. Here, $R_{\rm AVX}$ and $R_{\rm DP}$
    are, respectively, an absolute force calculated with our
    implementation and that obtained by performing all the
    calculations in double-precision without referring to the look-up
    table. The separations of particle pairs are distributed uniformly
    in $\ln(r)$ in the range of $5 \times 10^{-3} < r/r_{\rm cut} <
    1$.}  \label{fig:s2_error}
\end{figure}

Figure~\ref{fig:s2_error} shows a functional form of $R(r,\epsilon)$
(solid curve) and $f(r)$ (dashed curve) in the top panel and relative
errors of forces including both PP and PM parts, i.e. $R(r,\epsilon)$,
in the bottom panel as a function of $r/r_{\rm cut}$. In
Figure~\ref{fig:s2_error}, we can see that the relative errors are
less than $10^{-3}$, which are sufficiently accurate for cosmological
$N$-body simulations.

\section{Performance}
\label{sec:performance}

In this section, we present the performance of our implementation of
the collisionless $N$-body simulation using the AVX instructions
(hereafter AVX-accelerated implementation). For the measurement of the
performance, we use an Intel Core i7--2600 processor with 8MB cache
memory and a frequency of $3.40$ GHz, which contains four processor
cores. In measuring the performance, Intel Turbo Boost Technology is
disabled, and Intel Hyper-Threading Technology (HTT) is enabled. A
compiler which we adopt is {\tt GCC 4.4.5}, with options {\tt -O3
  -ffast-math -funroll-loops}. To see the advantage of the AVX
instructions relative to the SSE instructions, we also develop the
implementations with the SSE instructions rather than the AVX
instructions both for Newton's force and arbitrarily-shaped force
(SSE-accelerated implementation).

\subsection{Newton's force}

\begin{figure*}
  \begin{center}
    \includegraphics[scale=1.8]{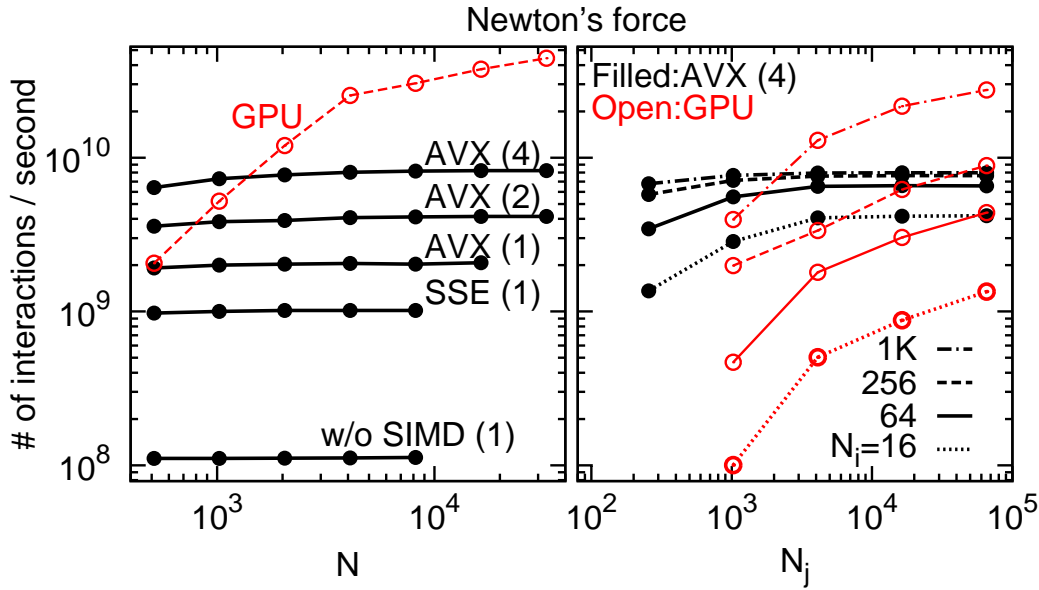}
  \end{center}
  \caption{(Left) Performance of the AVX- and SSE-accelerated
    implementation for Newton's force as a function of the number of
    particles (solid lines). The numbers in parentheses refer to the
    number of threads adopted. Dashed curve shows the performance of a
    GPU (NVIDIA GeForce GTX 580) attached to a host computer equipped
    with an Intel Core i7--2600K processor. (Right) Performance of the
    calculations of forces and potentials for a various set of the
    number of $i$- and $j$-particles, $N_i$ and $N_j$,
    respectively. Black and red curves show the performance with the
    AVX-accelerated implementation on four threads and that of
    GPU-accelerated implementation.} \label{fig:newton_pfm}
\end{figure*}

First, we show the performance of our implementation for Newton's
force. The performance is measured by executing the direct
particle-particle calculation of the Plummer model with the number of
particles from 0.5K to 32K. The left panel of
Figure~\ref{fig:newton_pfm} depicts the performances of the AVX- and
SSE-accelerated implementations. For comparison, we also show the
performance of an implementation without any explicit use of SIMD
instructions (labeled as ``w/o SIMD'' in the left panel of
Figure~\ref{fig:newton_pfm}). The numbers of interactions per second
are $2 \times 10^9$ in the case of the AVX-accelerated implementation
with a single thread, which corresponds to $75$~GFLOPS, where the
number of floating-point operations for the computation of force and
potential for one pair of particles is counted to be $38$. The
performances of the SSE- and AVX-accelerated implementations with a
single thread are higher than those without SIMD instructions by $10$
and $20$ times, respectively, and higher than those expected from the
degree of concurrency of the SSE and AVX instructions for
single-precision floating-point number (4 and 8, respectively). This
is because a very fast instruction of approximate inverse-square-root
is not used in the ``w/o SIMD'' implementation. On the other hand, the
performance with the AVX-accelerated implementation is higher than
that of the SSE-accelerated implementation roughly by a factor of two
as expected.

Furthermore, in the left panel of Figure~\ref{fig:newton_pfm}, we show
the performance of a GPU-accelerated $N$-body code based on the direct
particle-particle method implemented using the CUDA language, where
the GPU board is NVIDIA GeForce GTX 580 connected through the
PCI-Express Gen2 x16 link. The GPU-accelerated $N$-body code computes
the forces and potentials of the particles using GPUs, and integrate
the equations of the motion of the particles on a CPU. Thus, the
communication of the particle data between the main memory of the host
machine and the device memory on the GPU boards is required, and can
hamper the total efficiency of the code. Of course, if all the
calculations are performed on GPUs, we might not suffer from such
overhead. However, the performance of such implementation cannot be
fairly compared with those of the AVX- and SSE-accelerated
implementations, because the communication of the particle data is
inevitable when we perform $N$-body simulations with multiple GPUs or
with multiple nodes equipped with GPUs, regardless of the $N$-body
solvers such as Tree and TreePM methods.

The performances of the AVX- and SSE-accelerated implementations are
almost independent of the total number of particles, $N$. On the other
hand, the performance of the GPU-accelerated implementation strongly
depends on the number of particles $N$, due to the non-negligible
overhead caused by the particle data communication. For $N=0.5$K, the
performance of the GPU-accelerated implementation is only 5\% of that
for $N=32$K. Thus, for small $N$ ($0.5$K and $1$K), the performance of
the AVX-accelerated implementation with four threads is higher than
that with GPU-accelerated implementation, although, for large $N$
($4{\rm K}$--$32{\rm K}$), the performance of the GPU-accelerated
implementation is higher than that of the AVX-accelerated
implementation. These features can be explained by the communication
overhead in the GPU-accelerated implementation.

So far, we see the performance of our code in the case that both the
numbers of $i$- and $j$-particles ($N_i$ and $N_j$, respectively) are
the same and equal to $N$. However, in actual computations of forces
in collisionless $N$-body simulations based on various $N$-body
solvers such as PPPM, Tree, and TreePM methods, the numbers of $i$-
and $j$-particles $N_i$ and $N_j$ are much smaller than the total
number of particles $N$. In the Tree method modified for the effective
force with external hardwares or softwares as described in
\citet{Makino91}, for example, $N_i$ is the number of particles, for
which a tree traverse is performed simultaneously and the resultant
interaction list (size $N_j$) is shared, and typically around
$10$--$1000$. Furthermore, if one adopts the individual timestep
algorithm, the number of $i$-particles $N_i$ gets even smaller. The
number of $j$-particles $N_j$ is also decreased in Tree and TreePM
methods.  Therefore, we show the performance for typical $N_i$ and
$N_j$ in the realistic situations of typical collisionless $N$-body
simulations.

The right panel of Figure~\ref{fig:newton_pfm} shows the performance
of the AVX-accelerated implementation using four threads with four
processor cores (black lines) and that of the GPU-accelerated one (red
lines) for various set of $N_i$ and $N_j$. It can be seen that the
obtained performance gets lower for the smaller $N_i$ and $N_j$,
regardless of the implementations. For the AVX- and SSE-accelerated
implementations, this feature is due to the overhead of storing the
particle data into the structures \verb|Ipdata| and \verb|Jpdata|
shown in List~\ref{list:structures}. The amount of the overhead of
storing $i$- and $j$-particles are proportional to $N_i$ and $N_j$,
respectively, and the computational cost is proportional to $N_i
N_j$. Keeping this in mind the low performance with $N_i=16$ compared
with those with $N_i \ge 64$ can be ascribed to the overhead of
storing $j$-particles to the structure \verb|Jpdata|. For the
GPU-accelerated implementation, the overhead originates from the
transfer of the particle data to the memory on GPUs. It can be seen
that the performance of the AVX-accelerated implementation has rather
mild dependence on $N_i$ and $N_j$, while that of the GPU-accelerated
one relatively strongly depends on $N_j$. Such difference reflects the
fact that the bandwidths and latency of the communication between GPUs
and CPUs are rather poor compared with those of memory access between
CPUs and main memory. Thus, the performance of the GPU-accelerated
implementation is apparently superior to the AVX-accelerated one only
when both of $N_i$ and $N_j$ are sufficiently large (say, $N_i>1$K and
$N_j>4$K).

At the end of this section, we apply our AVX-accelerated
implementation to Barnes-Hut Tree method \citep{Barnes86}, and measure
its performance. Our tree code is based on the PP part of TreePM code
implemented by \cite{Yoshikawa05} and \cite{Fukushige05}, in which
they accelerated the calculations of the gravitational forces of the
$S2$-profile using GRAPE-5 and GRAPE-6A systems under the periodic
boundary condition. We modify the tree code such that it can compute
the Newton's force under the vacuum boundary condition. Since both of
GRAPE-6A systems and Phantom-GRAPE library support the same APIs, we
can easily utilize the capability of Phantom-GRAPE by simply
exchanging the software library.

\begin{figure*}[htbp]
  \leavevmode
  \begin{center}
    \includegraphics[scale=1.8]{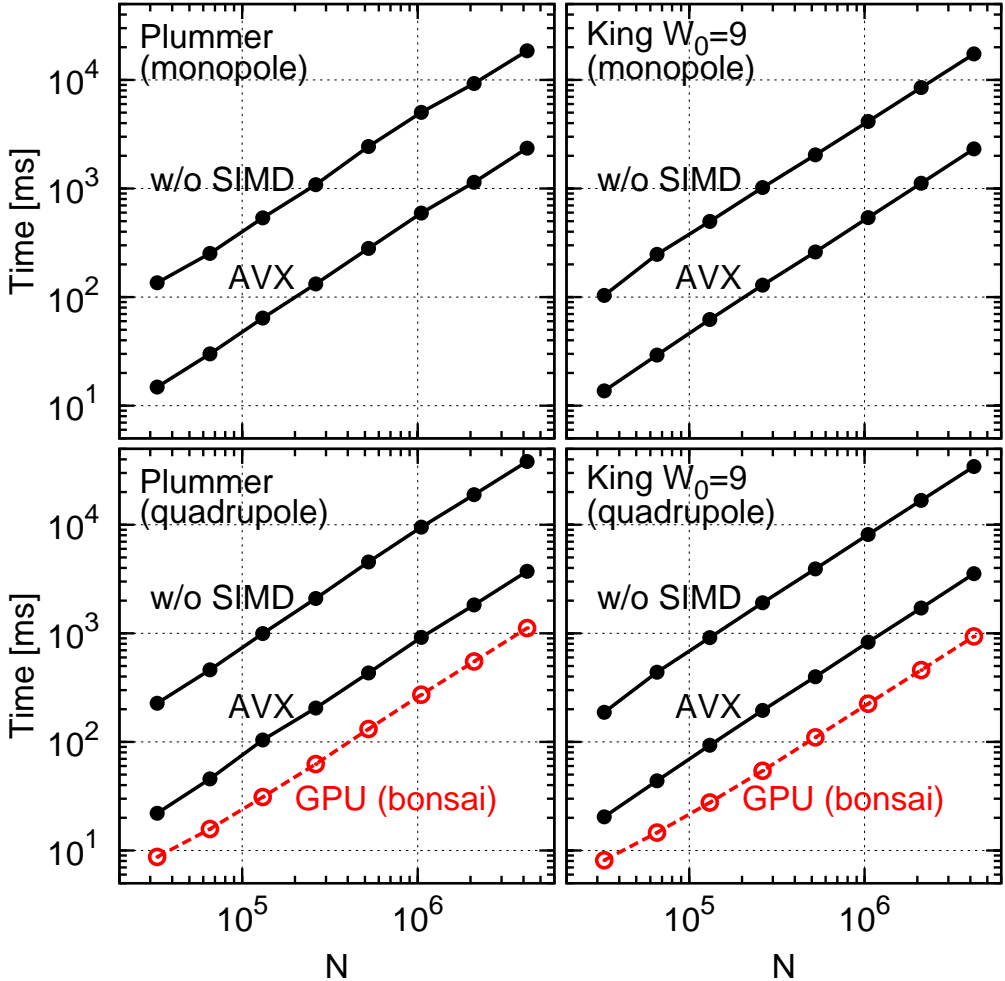}
  \end{center}
  \caption{Wall clock Time for calculating forces and potentials of all
    the particles in the Plummer model (left panels) and the King model
    (right panels) using Barnes-Hut tree method. The performance for
    codes which utilize only the monopole moment of the particle
    distribution in each tree nodes (upper panels) and ones which
    utilize monopole and quadrupole moments (lower panels) are
    shown. The opening parameter is set to $\theta = 0.75$. The
    performances of ``w/o SIMD'' and AVX-accelerated implementation are
    measured out with implementation using eight threads on an Intel
    Core i7--2600 processor, and that of the {\tt bonsai} code
    \citep{Bedorf12}, are on NVIDIA GeForce GTX 580.}
    \label{fig:treeNW}
\end{figure*}

Using the tree code described above, we calculate gravitational forces
and potentials of all the particles in a Plummer model and a King
model with the dimensionless central potential depth $W_0=9$. We
measure the performance on an Intel Core i7--2600 processor. For the
comparison with other codes, we also measure the performance of the
same code but without any explicit use of SIMD instructions, and the
publicly available code {\tt bonsai} \citep{Bedorf12}, which is a
GPU-accelerated $N$-body code based on the tree method. The
performance of the {\tt bonsai} code is measured on a system with
NVIDIA GeForce GTX 580. Since the {\tt bonsai} code utilizes the
quadrupole moments of the particle distribution in each tree node as
well as the monopole moments in the force calculations, for a fair
comparison of the performance with the {\tt bonsai} code, we give our
tree code a capability to use the quadrupole moments in each tree
node, although the original code uses only the monopole moments. We
represent these multipole moments as pseudo-particles, using
pseudo-particle multipole method
\citep{Kawai01}. Figure~\ref{fig:treeNW} shows the wall clock time to
compute gravitational forces and potentials for each tree code. We
show the both results with the code which uses the quadrupole moments
(lower panels) and the one which uses only the monopole moments (upper
panels). Note that the wall clock time includes the time for tree
construction, tree traverse and calculations of forces and potentials
but we exclude the time to integrate orbital motion of particles. As
expected, the wall clock time with the AVX-accelerated implementation
is roughly 10 times shorter than those without any explicit use of
SIMD instructions, owing to parallelism to calculate forces and
potentials. The wall clock time with the AVX-accelerated
implementation is about only three times longer than those with {\tt
  bonsai}, despite that theoretical peak performance of Intel Core
i7--2600 ($220$~GFLOPS) is lower than that of NVIDIA GeForce GTX 580
($1600$~GFLOPS) by a factor of 7.3 in single-precision. We expect that
the performance of our AVX-accelerated implementation could be close
to that of the {\tt bonsai} in the following situations. When we adopt
individual timestep algorithm, the number of $i$-particles is
effectively decreased, and a part of GPU cores becomes inactive. Thus,
the performance of GPU-accelerated implementation would be degraded
more rapidly than that of our AVX-accelerated
implementation. Furthermore, when we use GPU-accelerated
implementation on massively parallel environments, the communication
between CPUs and GPUs is inevitable, which also degrades the
performance of GPU-accelerated implementation.

\subsection{Force with an arbitrary shape}

\begin{figure*}
  \begin{center}
    \includegraphics[scale=1.8]{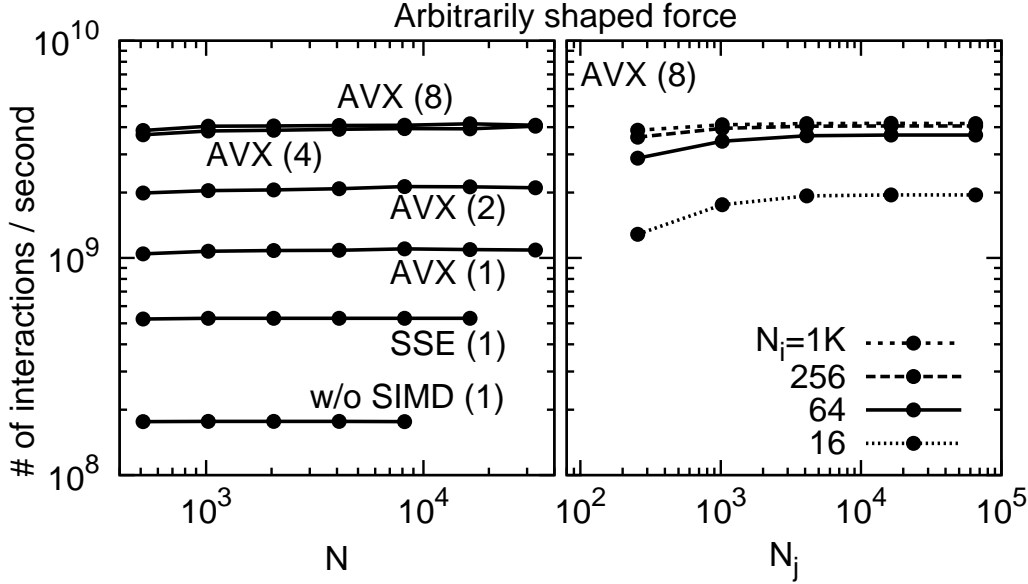}
  \end{center}
  \caption{(Left) Performance of calculations of arbitrarily shaped
    forces as a function of the number of particles, $N$. The vertical
    axis shows the number of interactions calculated per second. The
    numbers in parentheses beside indicate the number of
    threads. (Right) Performance of the AVX-accelerated implementation
    for the calculations of arbitrarily shaped forces for a various
    set of the number of $i$- and $j$-particles, $N_i$ and $N_j$,
    respectively. The forces are computed with the AVX instructions on
    eight threads.} \label{fig:s2_pfm}
\end{figure*}

The left panel of Figure~\ref{fig:s2_pfm} shows the performance of our
implementation to calculate forces with an arbitrary force shape
accelerated with the AVX and SSE instructions. For the comparison, we
also plot the performance of an implementation without any explicit
use of the SIMD instructions. The numbers of exponent and fraction
bits used to referring the look-up table are set to $E=4$ and $F=6$,
respectively. The performance of the AVX-accelerated implementation
with a single thread is $2$ and $6$ times higher than that of the
SSE-accelerated one and the one without any SIMD instructions,
respectively. These forces with the use of the AVX instructions are
lower than those expected from the degree of concurrency of their SIMD
operations, $8$, mainly because the reference of a look-up table is
not carried out in a SIMD manner. The performance with multi-thread
parallelization is almost proportional to the number of threads up to
four threads. If the HTT is activated, the performance with eight
threads is higher than that with four threads by a few percent.

The right panel of Figure~\ref{fig:s2_pfm} shows the performance of
the AVX-accelerated implementation with eight threads for a various
set of $N_i$ and $N_j$. For $N_i \ge 64$, the performance is almost
independent of $N_i$ and $N_j$, and for $N_i = 16$ it is about half
the performance with $N_i \ge 64$. This is again due to the overhead
of copying $j$-particle data to the structure \verb|Jpdata|, as is the
case in the calculation of Newton's force. Such weak dependence of the
performance on $N_i$ and $N_j$ are also preferable for the
calculations of the forces in the PPPM and TreePM methods especially
with the individual timestep scheme.

\section{Summary}
\label{sec:summary}

Using the AVX instructions, the new SIMD instructions of x86
processors, we develop a numerical library to accelerate the
calculations of Newton's forces and arbitrarily shaped forces for
$N$-body simulations. We implement the library by means of
inline-assembly embedded in C-language with GCC extensions, which
enables us to manually control the assignment of the YMM registers to
computational data, and extract the full capability of a CPU core. In
computing arbitrarily shaped forces, we refer to a look-up table,
which is constructed with a novel scheme so that the binning is
optimized to ensure good numerical accuracy of the computed forces
while its size is kept small enough to avoid cache misses.

The performance of the version for Newton's forces reaches $2 \times
10^9$ interactions per second with a single thread, which is about $2$
times and $20$ times higher than those of the implementation with the
SSE instructions and without any explicit use of SIMD instructions,
respectively. The use of the fast inverse-square-root instruction is a
key ingredient of the improvement of the performance in the
implementation with the SSE and AVX instructions. The performance of
the version for arbitrarily shaped forces is $2$ and $6$ times higher
than those implemented with the SSE instructions and without any
explicit use of the SIMD instructions. Furthermore, our implementation
supports the thread parallelization on a multi-core processor with the
{\tt OpenMP} programming interface, and has a good scalability
regardless of the number of particles.

While the performance of our implementation using the AVX instructions
is moderate compared with that of the GPU-accelerated implementation,
the most remarkable advantage of our implementation is the fact that
the performance has much weaker dependence on the numbers of $i$- and
$j$-particles than that of the GPU-accelerated implementation. This
feature is also the case for the calculation of the arbitrarily shaped
force, and can be explained by the relatively large overhead of the
data transfer between GPUs and main memory of their host computers. In
actual calculations of forces with popular $N$-body solvers such as
the Tree-method and the TreePM-method combined with the individual
timestep scheme, the numbers of $i$- and $j$-particles cannot be
always large enough to extract the full capability of GPUs. In that
sense, our implementation is more suitable in accelerating the
calculations of forces using the Tree- and TreePM-methods.

Another advantage of our implementation is its portability. With this
library, we can carry out collisionless $N$-body simulations with a
good performance even on supercomputer systems without any GPU-based
accelerators. Note that massively parallel systems with GPU-based
accelerators, at least currently, are not ubiquitous. Even on
processors other than the x86 architecture, most of them supports
similar SIMD instruction sets (e.g. Vector Multimedia Extension on IBM
Power series, and HPC-ACE on SPARC64 VIIIfx, etc.) Our library can be
ported to these processors with some acceptable efforts.

Finally let us mention the possible future improvement of our
implementation. Fused Multiply-Add (FMA) instructions which have
already been implemented in the ``Bulldozer'' CPU family by AMD
Corporation, and is scheduled to be introduced in the ``Haswell''
processor by Intel Corporation in 2013. The use of the FMA
instructions will improve the performance and accuracy of the
calculations of forces to some extent.  The numerical library
``Phantom-GRAPE'' developed in this work is publicly available at
{\ttfamily http://code.google.com/p/phantom-grape/}.

\section*{Acknowledgment}

We thank Dr. Takayuki Saitoh for valuable comments on this work.
A. Tanikawa thanks Yohei Miki and Go Ogiya for fruitful discussion on
GPU. Numerical simulations have been performed with computational
facilities at the Center for Computational Sciences in University of
Tsukuba. This work was supported by Scientific Research for
Challenging Exploratory Research (21654026), Grant-in-Aid for Young
Scientists (start-up: 21840015), the FIRST project based on the
Grants-in-Aid for Specially Promoted Research by MEXT (16002003), and
Grant-in-Aid for Scientific Research (S) by JSPS
(20224002). K. Nitadori and T. Okamoto acknowledge financial support
by MEXT HPCI STRATEGIC PROGRAM.

\end{document}